\documentclass[11pt,a4paper,total={6in, 8in}]{article}
\usepackage[T1]{fontenc}
\usepackage{graphicx,amssymb,amsmath,color}
\usepackage[linkcolor={blue},citecolor={red},colorlinks=true]{hyperref}
\usepackage{cite}
\usepackage{relsize}
\usepackage{enumerate}
\usepackage[margin=2cm]{geometry}
\usepackage{slashed}
\usepackage{multirow}
\usepackage[normalem]{ulem}
\usepackage[dvipsnames, x11names]{xcolor}
\usepackage[compat=1.0.0]{tikz-feynman}
\usepackage{xcolor}
\usepackage{mathtools}
\usepackage{cancel}
\usepackage{soul}
\usepackage{amsmath}
\usepackage{comment}

\def\beq{\begin{equation}}
\def\eeq{\end{equation}}
\def\bea{\begin{eqnarray}}
\def\eea{\end{eqnarray}}

\def\bit{\begin{itemize}}
\def\eit{\end{itemize}}

\def\sss{\scriptscriptstyle}

\def\baa{\begin{array}}
\def\eaa{\end{array}}

\def\d{\partial}

\def\simgt{\mathrel{\lower2.5pt\vbox{\lineskip=0pt\baselineskip=0pt
           \hbox{$>$}\hbox{$\sim$}}}}
\def\simlt{\mathrel{\lower2.5pt\vbox{\lineskip=0pt\baselineskip=0pt
           \hbox{$<$}\hbox{$\sim$}}}}

\def\bfc{\begin{figure}\begin{center}}
\def\efc{\end{center}\end{figure}}
\def\nn{\nonumber\\}

\definecolor{chromeyellow}{rgb}{1.0, 0.65, 0.0}
\definecolor{darkcoral}{rgb}{0.8, 0.36, 0.27}
\definecolor{cadmiumgreen}{rgb}{0.0, 0.42, 0.24}

\begin{document}

\begin{flushright}
\hspace{3cm} 
\\

\end{flushright}
\vspace{.6cm}
\begin{center}

\hspace{-0.4cm}{\Large \bf 
Dynamical evolution of the pressure on the bubble wall  \\}

\begin{center}
 Benoit Laurent$^{\dagger\,1}$ and Miguel Vanvlasselaer$^{\ddagger \, 2}$ \\
\vskip0.4cm

{\it $^1$Perimeter Institute for Theoretical Physics, Waterloo, Ontario N2L 2Y5, Canada} \par
\vskip0.2cm
{\it $^2$Departament de Física Quàntica i Astrofísica and Institut de Ciències del Cosmos (ICC), Universitat de Barcelona, Martí i Franquès 1, ES-08028, Barcelona, Spain.}
\end{center}

\end{center}

\bigskip \bigskip \bigskip

\centerline{\bf Abstract} 
\begin{quote}
First-order phase transitions in the early Universe are pivotal for gravitational wave production, baryogenesis, and dark matter generation. A central question is whether bubble walls reach a subjouguet or ultra-relativistic velocity — a distinction governed by hydrodynamic obstruction, where plasma heating counteracts the vacuum pressure driving the wall. Traditional analyses assume steady-state fluid profiles, but these may fail during the wall’s acceleration phase.
We study the dynamical evolution of the pressure on the bubble wall in local thermal equilibrium (LTE), combining analytical approximations with numerical hydrodynamic simulations. Our results reveal that the heating wave’s formation time often exceeds the wall’s acceleration timescale, invalidating steady-state predictions near the Jouguet velocity. We derive a revised criterion for the maximal driving pressure, which separates deflagration/hybrid regimes from detonations/runaway walls. This criterion, validated by simulations, shows that hydrodynamic obstruction is less restrictive than steady state LTE predictions suggest.

\end{quote}

\vfill
\noindent\line(1,0){188}
{\scriptsize{ \\ E-mail:
\texttt{$^\dagger$\href{blaurent@perimeterinstitute.ca}{blaurent@perimeterinstitute.ca}},
$^\ddagger$\href{miguel.vanvlasselaer@ub,edu}{miguel.vanvlasselaer@ub.edu}}
}

\newpage
\hrule
\tableofcontents
\vskip .85cm
\hrule
\newpage

\section{Introduction}

Cosmological first-order phase transitions (FOPTs) have far-reaching phenomenological
consequences. The expansion and collision of the nucleated bubbles could result in a
stochastic background of gravitational waves (GWs)~\cite{Witten:1984rs,Kosowsky:1991ua,
Kosowsky:1992vn,Kamionkowski:1993fg,Hindmarsh:2013xza,Caprini:2015zlo,Caprini:2019egz}.
Renewed interest in FOPTs in the early Universe has been triggered by the recent approval
of the Laser Interferometer Space Antenna (LISA) project~\cite{LISA:2017pwj} and the
detection of GWs~\cite{LIGOScientific:2016aoc}. Besides LISA, many other space-based GW
detectors have been proposed, such as Big Bang Observer
(BBO)~\cite{Corbin:2005ny}, Deci-hertz Interferometer Gravitational wave Observatory
(DECIGO)~\cite{Kawamura:2011zz}, Taiji~\cite{Gong:2014mca}, and
TianQin~\cite{TianQin:2015yph}. Excitingly, such an SGWB has been reported by several Pulsar Timing Array
projects~\cite{NANOGrav:2023gor,EPTA:2023fyk,Reardon:2023gzh,Xu:2023wog} recently,
whose source however may prefer a supermassive black hole explanation.

FOPTs may also be relevant for producing primordial black
holes~\cite{Kodama:1982sf,Hawking:1982ga,Garriga:2015fdk,Deng:2017uwc,Gross:2021qgx,
Baker:2021nyl,Kawana:2021tde,Liu:2021svg,Jung:2021mku,Huang:2022him,Lewicki:2023ioy,
Gouttenoire:2023naa}, particle dark matter~\cite{Baker:2019ndr,Chway:2019kft,Chao:2020adk},
intergalactic magnetic fields~\cite{Vachaspati:1991nm,Olea-Romacho:2023rhh} and the observed baryon number \cite{Kuzmin:1985mm,Shaposhnikov:1987tw,Morrissey:2012db,Garbrecht:2018mrp,Shaposhnikov:1987tw, Cline:2020jre, Cline:2021dkf, Ellis:2022lft}.
All the above-mentioned processes depend crucially on the regime of the bubble wall
expansion, which can reach a terminal constant velocity, or can accelerate until bubble wall collision (runaway regime). For example, in recent years,
ultrarelativistic bubble walls (bubbles expanding with a Lorentz factor
$\gamma_w \equiv 1/\sqrt{1-\xi_w^2} \gg 1$, where $\xi_w$ is the wall velocity) have been shown to offer  novel channels
of baryogenesis~\cite{Cline:2020jre,Azatov:2021irb,Baldes:2021vyz,Huang:2022vkf,
Chun:2023ezg,Dichtl:2023xqd, Cataldi:2025nac, Cataldi:2024pgt, Vanvlasselaer:2026fay} and dark matter
production~\cite{Azatov:2021ifm,Azatov:2022tii,Borah:2022cdx,Baldes:2023fsp,
Giudice:2024tcp}. For the application of ultrarelativistic bubble walls, it is extremely
important to determine whether the bubble wall can enter into the ultrarelativistic regime
or not~\cite{Banerjee:2024qiu}. Furthermore, runaway and non-runaway walls could have
very different GW signals (see
Refs.~\cite{Caprini:2015zlo,Caprini:2018mtu,Caprini:2019egz,
LISACosmologyWorkingGroup:2022jok,Athron:2023xlk,Roshan:2024qnv} for reviews and see,
e.g., Refs.~\cite{Jinno:2019jhi,Lewicki:2022pdb} for studies on supercooled PTs; see also \cite{Lewicki:2024xan} for the impact of the uncertainties on the GW signal). Conversely, it is also well-known that the wall velocity has a strong impact on the Electroweak baryogenesis\cite{Shaposhnikov:1987tw, Cline:2020jre, Cline:2021dkf, Ellis:2022lft}  (see also \cite{Barni:2025ifb} for a review of the baryogenesis literature).

Determination of the regime of expansion and estimates of the terminal wall velocity are usually based on kinetic
theory~\cite{Dine:1992wr,Liu:1992tn,Moore:1995ua,Moore:1995si}, improvement of
it~\cite{Cline:2020jre,Cline:2021iff,Laurent:2022jrs}, or holographic methods for
strongly coupled theories~\cite{Bea:2021zsu,Baldes:2020kam,Bigazzi:2020avc,
Bigazzi:2021ucw,Bea:2022mfb,Li:2023xto,Wang:2023lam} (see Ref.~\cite{Kang:2024xqk}
for a quasiparticle (quasigluons) method). Typically, calculating the terminal wall
velocity is a very challenging task, and has only been performed for a limited number of
models~\cite{Liu:1992tn,Moore:1995si,Moore:1995ua,Dorsch:2018pat,Laurent:2022jrs,
Jiang:2022btc} (see \cite{Ekstedt:2024fyq} for an automatized code of the bubble wall velocity). In such computations, one needs to solve the scalar field equation of
motion (EoM) coupled to the Boltzmann equations describing the particles in the plasma.
Simplifications can be obtained by assuming local thermal equilibrium
(LTE)~\cite{Konstandin:2010dm,BarrosoMancha:2020fay,Balaji:2020yrx,Ai:2021kak,
Wang:2022txy,Ai:2023see,Krajewski:2023clt,Sanchez-Garitaonandia:2023zqz} or in the
ultrarelativistic regime~\cite{Bodeker:2009qy,Bodeker:2017cim,Hoche:2020ysm,
Azatov:2020ufh,Gouttenoire:2021kjv,Ai:2023suz,Azatov:2023xem}. In LTE, the wall
velocity can be determined with the help of an additional matching condition for the
hydrodynamic quantities on both sides of the
wall~\cite{Ai:2021kak,Ai:2023see}. The LTE approximation is shown to work well for
strongly coupled theories~\cite{Sanchez-Garitaonandia:2023zqz} whose FOPT-related
phenomenology has been studied in,
e.g., Refs.~\cite{Schwaller:2015tja,Aoki:2017aws,Helmboldt:2019pan,Agashe:2019lhy,
Bigazzi:2020avc,Halverson:2020xpg,Huang:2020crf,Ares:2020lbt,Kang:2021epo,
Reichert:2021cvs,Morgante:2022zvc,He:2022amv,Fujikura:2023fbi,Pasechnik:2023hwv, Cline:2025bwe}, moreover, it has recently been applied to the inverse phase transitions in \cite{Barni:2025gnm}.

The expanding phase boundary (in the case of a hybrid and a deflagration) induces a hot compression wave in front of the wall. This induces a heating of the phase boundary implying the weakening of the transition\cite{Witten:1984rs}. 
This heating of the wall becomes more and more efficient as the wall velocity is increased
toward the Jouguet velocity~\cite{Ai:2023see,Ai:2024shx}, velocity at which the compression wave is very thin and the heating is maximal. For larger velocities, the
compression wave disappears and the heating effect becomes inefficient. This pattern of
increasingly relevant heating until Jouguet velocity followed by a vanishing of the effect
leads to a peak (a point of maximal resistance) in the effective pressure felt by the
scalar wall, which today goes under the name of \emph{hydrodynamical
obstruction}~\cite{Konstandin:2010dm}.

The temperature at the phase boundary (behind the compression wave) can be directly obtained from the fluid profile around the bubble. In the asymptotic regime, once the bubble wall reaches a terminal constant velocity, the fluid profile admits an attractor solution — a steady state expressible through the self-similar variable  $\xi \equiv r/t$. Crucially, this ``Steady State'' (SS) attractor is not reached instantaneously, but rather emerges as the endpoint of a finite-time evolution (see for
example~\cite{Jinno:2022mie} for a 3d approach to ``steady state'' study). More
specifically, we will see that convergence to the SS typically takes a time
$t \sim \text{few}\, R_0$ (where  $R_0$  is the bubble radius at nucleation)  for detonations and deflagrations, growing to
$t \sim \text{several tens of}\, R_0$ for hybrids close to the Jouguet velocity (beyond
which the compression wave just disappears). The SS of deflagrations and hybrids fluid
profile both present a hot compression wave in front of the scalar wall, implying that
the effective temperature ``seen'' by the scalar wall is that of the compression wave
$T_+ > T_{n}$~\cite{Ai:2021kak}, larger than the nucleation temperature.

  To go further, we now define 
\begin{subequations}
\begin{align}
   \alpha_p(T) &\equiv \frac{4\Delta V_{\rm eff}(T)}{3w_+(T)}\\ 
   \Psi(T) &\equiv \frac{w_-(T)}{w_+(T)}\approx\frac{g_-^\star}{g_+^\star} \,  ,
\end{align}
\end{subequations}
where the $+/-$ subscripts refer to the quantity evaluated right ahead/behind the wall. Furthermore, $V_{\rm eff}$ is the effective potential, $w$ the enthalpy density and $g^\star$ the effective number of degrees of freedom.
Importantly, using the \emph{steady state} LTE approach and the high-$T$ expanded potential used in this paper (see Appendix \ref{sec:potential}), one can show that any phase transition with strength $\alpha_p$ smaller than
\begin{align}\label{LTE:criterion}
    \alpha_{p,{\rm max}}^{\rm SS}\approx\frac{10}{9}\sqrt{\frac{2}{3}}(1-\Psi)^{3/2}
\end{align}
would lead to a terminal velocity \emph{smaller} than the Jouguet velocity, while any transition with larger $\alpha_p$ would lead to a runaway behaviour. We refer the reader to Appendix \ref{app:alphaMax} for details about this calculation.

However, the result presented above and most of the past studies relied on the peculiar assumption that the fluid profile around the bubble wall instantaneously tracks the SS. As we emphasized above, the SS is only reached after a finite evolution timescale, which can be even larger than the timescale for the bubble wall to reach the Jouguet velocity. This intuition invites a deeper inspection of the assumption of instantaneous SS and a reassessment of the impact of hydrodynamical obstruction.

In fact, the authors of \cite{Krajewski:2023clt,Krajewski:2024zxg, Krajewski:2024gma} developed a real time evolution code \cite{Krajewski:2023clt} of the coupled system wall-plasma which solves the hydrodynamical equations and the equation of motion of the bubble wall in 1+1 dimensions (assuming spherical symmetry of the bubble). The effect of the phenomenological friction induced by the entropy production was studied in~\cite{Krajewski:2024gma}.  These early studies confirmed that, within the context of the extended standard model (xSM), the early dynamical evolution could partially evade hydrodynamical obstruction. The aim of this paper is twofold: first we aim to extend these results in a model-independent way to a broad range of values of $\Psi$, and, second, we will provide a simple quantitative criterion to determine if hydrodynamic obstruction occurs.

\begin{figure}[h]
 \centering  \includegraphics[width=0.49\textwidth]{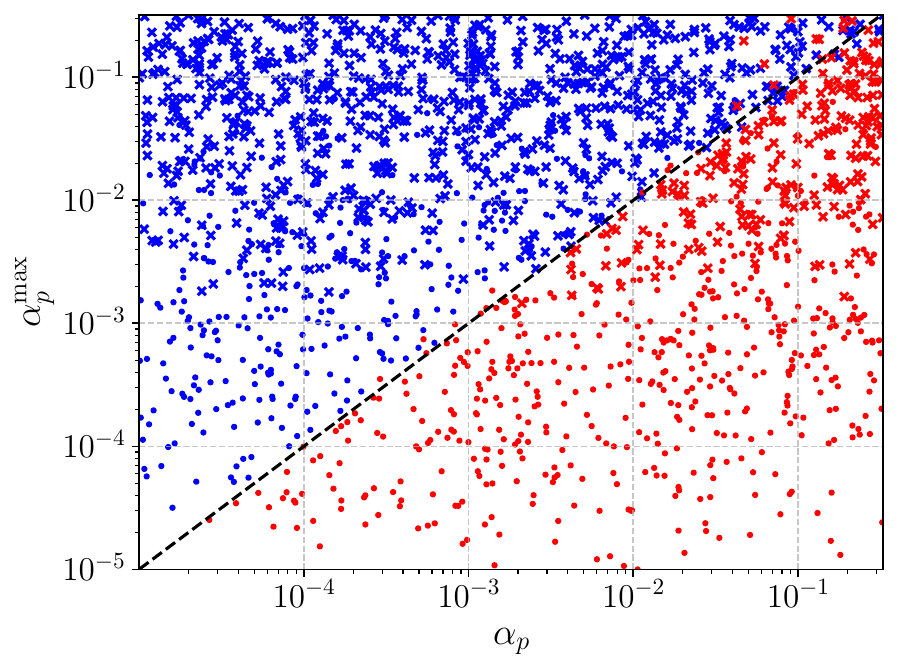}
 \includegraphics[width=0.49\textwidth]{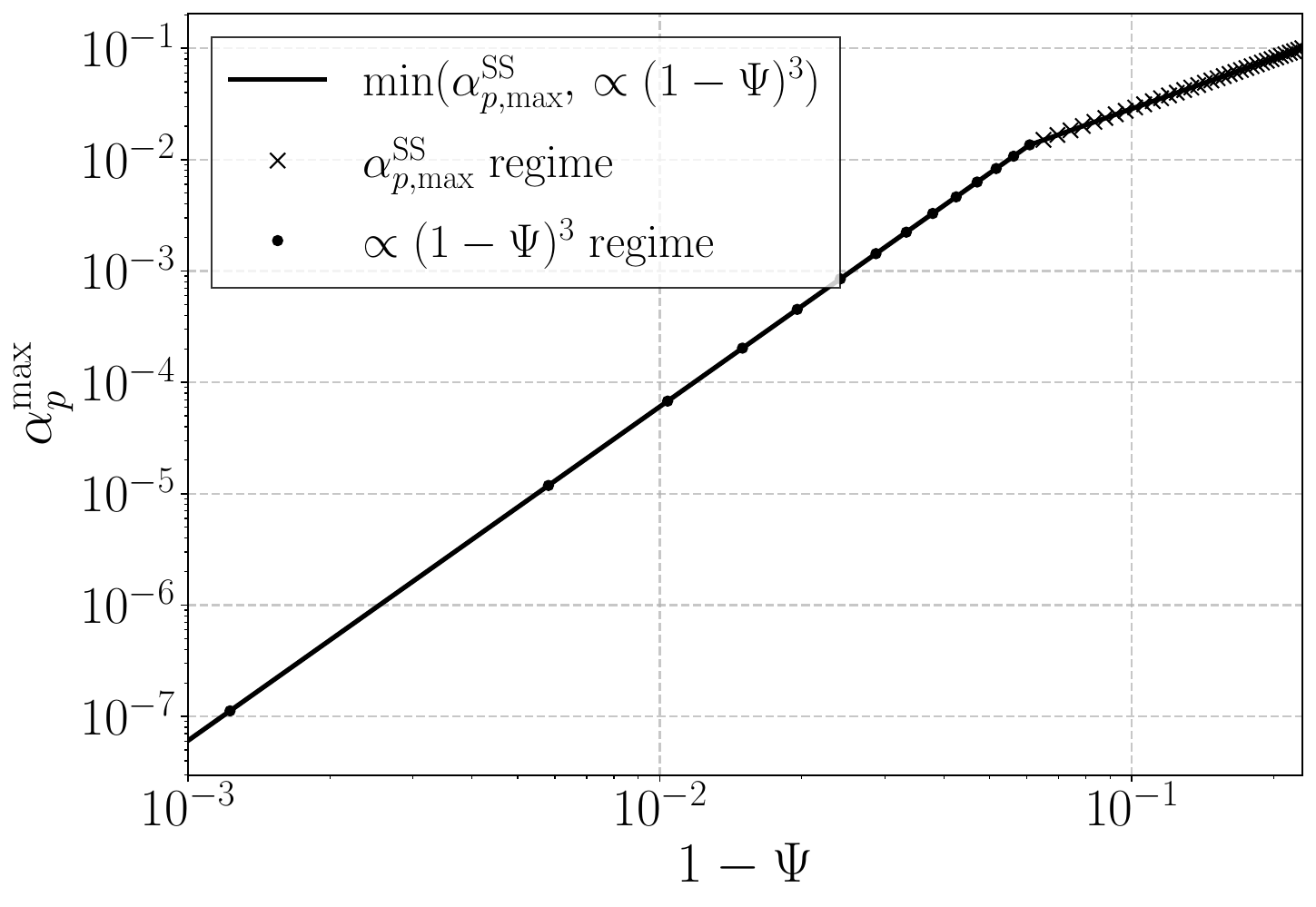}\
 \caption{\textbf{Left panel}: Scatter plot of the value of $\alpha_p^{\rm max}$ defined in Eq.~(\ref{LTE:criterion}) as a function of $\alpha_p$. The red markers are runaway solutions, while the blue markers are deflagration or hybrid solutions that have reached a terminal velocity. For the crosses (dots) $\alpha_{p}^{\rm max}$ is computed using the first (second) term in the min of Eq.~\eqref{eq:main_result}. The details of this scan are described in Section \ref{sec:results}. $\alpha_p^{\rm max}$ is computed from Eq.~\eqref{eq:runawayCriterion} with the optimal fit given by $\xi_w^{\rm max}\simeq 0.81$. \textbf{Right panel}: Definition of $\alpha_p^{\max}$ as a function 
of $\Psi$ used on the left panel, where we have set $T_n = \phi_0$, $g_+^*=106.75$ and $S_3/T_n = 140$.  The definition of the $\alpha_p^{\rm max}$ used on the left panel is given by the black curve, 
the physical upper bound presented in Eq.\eqref{eq:main_result}. The crosses (on both panels) are computed with $\alpha^{\rm max}_p = \alpha^{\rm static}_{p, \rm max}$ while dots are computed with $\alpha^{\rm max}_p = \alpha^{}_{p, \rm max}$ obtained from Eq.~\eqref{eq:main_result}. }
 \label{fig:scatter_plot}
 \end{figure}

\paragraph{Results}
The main result of this paper is to provide a simple criterion to determine in which region of the parameter space the hydrodynamical obstruction develops fast enough to prevent the formation of a detonation or a runaway regime. Contrarily to the criterion obtained from the steady state LTE approach in Eq.~\eqref{LTE:criterion}, we obtain the following criterion (computed from the dynamical evolution equations)
\begin{align}
\label{eq:main_result}
 \alpha_{p}^{\rm max} \approx \text{min}\bigg[  \alpha_{p,{\rm max}}^{\rm SS},\, \,  60.64\left(\frac{1-\Psi}{\phi_0/T_n}\right)^3\left(\frac{\nu}{4}\right)\left(\frac{S_3/T_n}{140}\right)\left(\frac{g_+^*}{106.75}\right)^{1/2}\left(\frac{\gamma_w^{\rm max}\xi_w^{\rm max}}{1.381}\right)^3 \bigg] \, , 
\end{align}
with $\phi_0$ the vacuum expectation value (VEV) acquired by the scalar field triggering the phase transition. Eq.~(\ref{eq:main_result}) constitutes the core result of this paper. The second term in the min is obtained from a combination of fit on the data and analytical expansion, as this will be explained in Section \ref{sec:dyn}.
In Fig.~\ref{fig:scatter_plot} we present a scatter plot of the regime of expansion of a bubble, deflagration (or hybrid) in blue and detonations (or runaway) in red, in the plane $\alpha_p^{\rm max}$ versus $\alpha_p$. For the crosses (dots) $\alpha_{p}^{\rm max}$ is computed using the first (second) term in the min of Eq.~\eqref{eq:main_result}. This criterion is has been numerical verified in the range of parameter space
\begin{align}
    \alpha_p&\in[10^{-5}, 0.33], \qquad \qquad 
    1-\Psi \in [0.01, 0.5], \qquad \qquad 
    \phi_0/T_n\in [0.5,2].\nonumber
\end{align}
with a precision of around 98\% (See Section \ref{sec:dyn} for all the details of the numerical analysis).

\paragraph{Organisation}
The remainder of this paper is organised as follows: in Section \ref{sec:early_acc}, we review the physics of bubble acceleration analytically and numerically. In Section \ref{sec:LTE}, we remind the reader of the steady state LTE picture and present, using numerical simulations, the limitations of this approach.  In Section \ref{sec:dyn}, we present an analytical derivation of $\alpha_p^{\rm max}$ and a simulation of a coupled system of a scalar field and fluid which confirms the validity of the criterion. Finally, in Section \ref{sec:conclusion}, we conclude and discuss the limitations of our approach.

\section{Early bubble wall expansion}
\label{sec:early_acc}

In this first section, we remind concisely the physics of early bubble wall expansion. We will pay particular attention to the acceleration of the wall as a function of time. 
The evolution of the bubble wall is controlled by the equation of motion of the scalar field
\begin{align}
 \label{eq:eom}
  \Box\phi+\frac{\d V_{\rm eff}(\phi, T)}{\d\phi}=0,
 \end{align}
 where the potential $V_{\rm eff}(\phi, T)$ is the potential where the subscript eff stands for a generic loop-resummed and thermally corrected potential\footnote{The potential that we will use throughout this paper is presented in Eq.~\eqref{eq:bagPotential} of  Appendix \ref{sec:potential} and can be understood as a high-$T$ expansion of $V_{\rm eff}$.}. In this section, we specialize to a bubble expanding in vacuum (we remove by hand all the temperatures in the potential $V_{\rm eff}(\phi, T)$) and simulate this case numerically. 

If we assume that the bubble is formed out of a spherically symmetric scalar field $\phi(t,r)$, Eq.~\eqref{eq:eom} becomes
\begin{align}\label{eq:wallEOM}
 \text{(Spherical Bubble equation of motion):} \qquad\qquad     \ddot{\phi}=\phi''+\frac{2}{r}\phi'-\frac{\partial V_{\rm eff}}{\partial\phi},
\end{align}
where the dots and primes represent time and radial derivatives, respectively. This approach neglects the contribution on the pressure coming from the out-of-equilibrium particles crossing the wall.

\subsection{Initial acceleration}

\paragraph{Analytical derivation}

To derive an analytical result, we further assume that the field profile can be described  by an ansatz of the form
\begin{align}\label{eq:fieldAnsatz}
    \phi(t,r)\equiv\phi[\gamma_{\sss R}(t)(r-R(t))],
\end{align}
with $R(t)$ the time-dependent wall position and $\gamma_{\sss R}(t)=1/\sqrt{1-\dot{R}^2}$ its Lorentz factor. This is expected to be a valid approximation in the thin-wall limit, when $R\gg L$, with $L$ the wall width.

One can now multiply Eq.~(\ref{eq:wallEOM}) by $-\phi'$ and integrate with respect to $r$. Using the ansatz (\ref{eq:fieldAnsatz}), the left-hand side becomes
\begin{align}
    -\int\! dr\, \phi'\ddot{\phi}&=\gamma_{\sss R}^3\ddot{R}\int \! d(\gamma_{\sss R} r)\left(\frac{\partial\phi}{\partial(\gamma_{\sss R} r)}\right)^2
    \equiv \gamma_{\sss R}^3\ddot{R}\sigma,
\end{align}
where we have assumed $\phi'[-\gamma_{\sss R}(r-R)]=\phi'[\gamma_{\sss R}(r-R)]$ and have neglected all the odd terms in $r-R$, which vanish in the integration. We have also defined the surface tension via 
\begin{align}
  \sigma &\equiv \frac{1}{\gamma_{\sss R}}\int\! dr\, (\phi')^2 \equiv C \frac{\phi_0^2}{3L} \, ,
\end{align}
with $C$ an $\mathcal{O}(1)$ constant. For example, if $\phi$ follows a tanh profile, one easily finds $C=1$, which is the value we will adopt in what follows. Multiplying Eq.~(\ref{eq:wallEOM}) by $-\phi'$ and integrating the right-hand side of the field EoM then yields
\begin{align}
\label{eq:eomRHS}
    -\int\! dr\, \phi'\left(\phi''+\frac{2}{r}\phi'-\frac{\partial V_{\rm eff}}{\partial\phi}\right) &\approx -\frac{2}{R}\gamma_{\sss R}\int \! d(\gamma_{\sss R} r)\left(\frac{\partial\phi}{\partial(\gamma_{\sss R} r)}\right)^2+\int\!dr\,\phi'\frac{\partial V_{\rm eff}}{\partial\phi}\nn
    &=  -\frac{2}{R}\gamma_{\sss R}\sigma+\mathcal{P}_{\rm LTE}(t) \, ,
\end{align}
where we defined the pressure on the bubble wall, computed from the total potential, via
\begin{align}
\label{eq:pressure_LTE}
    \mathcal{P}_{\rm LTE}(t)&= \int\! dr\, \phi'\frac{\partial V_{\rm eff}}{\partial\phi} = \Delta V_{\rm eff} - \int\! dr\, T'\frac{\partial V_{\rm eff}}{\partial T} \, . 
\end{align}
As we already emphasized, this expression of the pressure neglects the contribution coming from the out-of-equilibrium particles crossing the wall.
In Eq.~\eqref{eq:eomRHS}, to obtain the first term, we have used the fact that, in the thin-wall limit, $\phi'$ is nonzero only in the close neighborhood of $r=R$, so it is justified to replace $1/r$ with $1/R$. We finally see that the wall position $R(t)$ follows the equation of motion
\begin{align}\label{eq:wallEOM1}
    \ddot{R}+\frac{2}{R}(1-\dot{R}^2) = \frac{\mathcal{P}_{\rm LTE}(t)}{\sigma}(1-\dot{R}^2)^{3/2} \, . 
\end{align}

The size of the critical bubble (corresponding to the bounce solution) can be obtained directly from Eq.~(\ref{eq:wallEOM1}) by setting $\ddot R=\dot R=0$, which yields
\begin{align}
   \text{(Critical radius)}\qquad \qquad R_0 = \frac{2\sigma}{\Delta V_{\rm eff}}.
\end{align}
Furthermore, it can be instructive to find an equation describing $\gamma_{\sss R}$ as a function of $R$. Using the identity $\ddot R=\partial_{\sss R}\gamma_{\sss R}/\gamma_{\sss R}^3$, one finds the simple equation
\begin{align}
\label{eq:eqoM}
    \partial_{\sss R}\gamma_{\sss R}+\frac{2}{R}\gamma_{\sss R} = \frac{\mathcal{P}_{\rm LTE}(R)}{\sigma}.
\end{align}
In the case where the bubble propagates in vacuum and $\mathcal{P}_{\rm LTE} = \Delta V_{\rm eff} = 2\sigma/R_0$ is a constant, this equation has the exact solution
\begin{align}\label{eq:vacuumSolution}
   \text{(Vacuum bubble expansion)} \qquad \qquad \gamma_{\sss R}^{\rm vac}(R) &= \frac{\mathcal{P}_{\rm LTE}R}{3\sigma}+\frac{4\sigma^2}{3\mathcal{P}_{\rm LTE}^2R^2} = \frac{2R}{3R_0}+\frac{R_0^2}{3R^2},
\end{align}
which satisfies the boundary condition $\gamma_{\sss R}^{\rm vac}(R_0)=1$.

An important consequence of this solution is that the bubble acceleration is quite fast. If the pressure is not suppressed by thermal or particle physics effects, it reaches the speed of sound when $R\approx 1.65R_0$. Thus, the plasma must also react and produce a compression wave on a short timescale for hydrodynamic obstruction to be efficient. If the formation timescale is too long, the boost factor quickly enters the linear growth regime, where $\gamma_{\sss R} \propto R/R_0$ and the wall becomes relativistic.

If a plasma surrounds the bubble, as the bubble accelerates, the pressure drops and in some cases falls to zero. When this happens, the bubble reaches a constant velocity, the \emph{terminal velocity}. In the opposite case, it runs away and keeps accelerating forever.

\paragraph{Dynamical simulation}
To confirm the validity of this effective description of the wall, we 
compare it to a full simulation of the early scalar field evolution in 
vacuum by solving the EoM~\eqref{eq:wallEOM}; see Section~\ref{sec:results} 
and Appendix~\ref{sec:numericalAlgorithm} for details.

\begin{figure}[t]
 \centering  \includegraphics[width=0.48\textwidth]{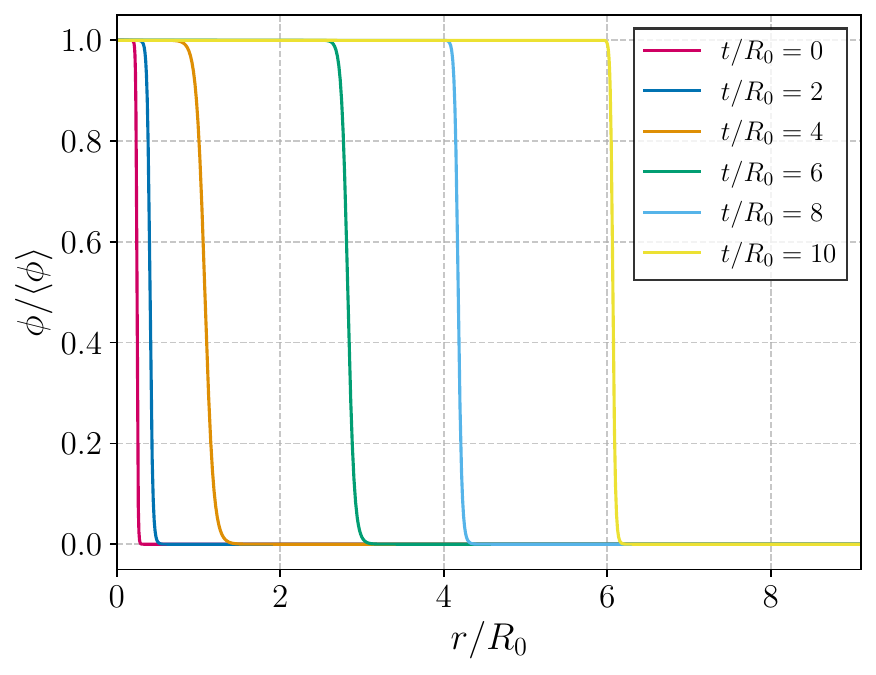}\hspace{0.03\textwidth} \includegraphics[width=0.48\textwidth]{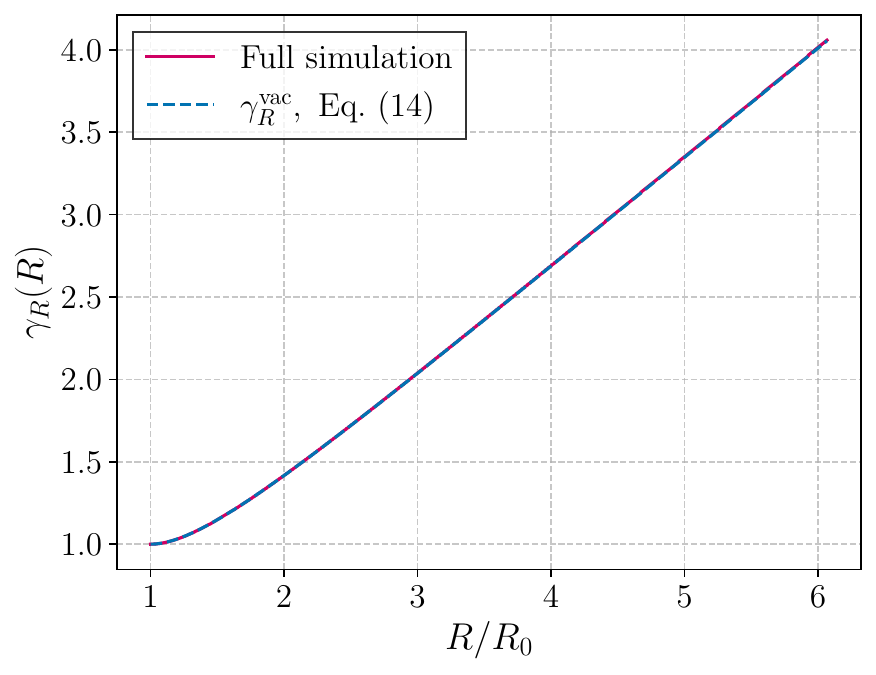}\\
 \caption{ {\textbf{Left panel}}: Early evolution of a scalar field profile of a $\mathcal{O}(3)$-symmetric bubble in vacuum. {\textbf{Right panel:}} Lorentz factor of the wall as a function of its position computed with the full dynamical simulation of the scalar field and with Eq.~(\ref{eq:vacuumSolution}).}
 \label{fig:early_exp}
 \end{figure}

The results are shown in Fig.~\ref{fig:early_exp}. The left panel shows 
the scalar field profile at several times after bubble nucleation. The 
right panel shows the Lorentz factor of the wall as a function of its 
position and compares it to the effective description Eq.~\eqref{eq:vacuumSolution}. 
The agreement between the two is excellent, validating our earlier result.

\subsection{Wall trajectory in the presence of a plasma}

As we have emphasized above, the pressure computed from the potential alone $\mathcal{P}_{\rm LTE}$ does not take into account the unavoidable out-of-equilibrium effects. These effects have been copiously discussed in the literature, namely by using two computations methods: in the first approach, the kinetic approach, one solves the coupled equations of motion (EoMs) for the background scalar field (condensate) and the plasma \,\cite{Liu:1992tn,Konstandin:2010dm,Huber:2011aa,Kozaczuk:2015owa,Dorsch:2018pat,Laurent:2020gpg,Friedlander:2020tnq,Balaji:2020yrx,Cline:2021iff,Ai:2021kak,Lewicki:2021pgr,Dorsch:2021nje,Jiang:2022btc,Laurent:2022jrs,Wang:2022txy,Ai:2023see,Krajewski:2023clt,Wang:2023kux,Dorsch:2023tss,Kang:2024xqk,Ai:2024shx,Krajewski:2024gma,Wang:2024wcs,Barni:2024lkj,Ekstedt:2024fyq,Ai:2024btx,Krajewski:2024xuz,Dorsch:2024jjl,Branchina:2025adj, Branchina:2025jou, DeCurtis:2022hlx, Krajewski:2026kcm} and, in the second approach, the ballistic computations\cite{Dine:1992wr,Bodeker:2009qy, Bodeker:2017cim,Liu:1992tn,BarrosoMancha:2020fay,Wang:2024wcs,Ai:2024btx, Hoche:2020ysm, Azatov:2020ufh,Gouttenoire:2021kjv,GarciaGarcia:2022yqb,Ai:2023suz,Azatov:2023xem,Baldes:2024wuz,Azatov:2024auq} the wall is modeled as a fixed background which receives kicks from particles flying through it. 

\paragraph{Ballistic approach}

When the wall is fast enough, particles pass through the wall so quickly that there is no time for them to collide with each other during their passage through the wall. We are thus justified in neglecting the interactions across nearby particles. Often, this also means that the bubble expansion is assumed to be in the so-called detonation mode, such that the fluid in front of the wall is not perturbed. In this approach, the friction simplifies to
\[
\mathcal{P}_{\text{kick}} =
\sum_{a,X} \int \frac{d^3 p}{(2\pi)^3 p^0} \, p^z \,
\Phi_{a \to X}(p)\, f_a(p)\, \Delta p^z_{a \to X}
\quad \text{(friction in the kick picture)} \, ,
\]
where $\Phi_{a \to X}(p)$ is the probability of the interaction $a \to X$ inside the wall, $\Delta p^z_{a \to X}$ is the exchange of momentum between the particles and the wall. The interpretation of this expression is thus straightforward: the pressure on the bubble is the sum over all the momentum that is lost by the plasma particles. 

Even though this approach is less generic than the kinetic approach, it lead in the past to the discovery of genuinely novel contributions to the pressure \cite{Bodeker:2017cim, Azatov:2020ufh}.

\paragraph{Kinetic approach}

By acting on the particles of the plasma, the wall induces a departure from thermal distributions, departure which is discarded by construction in the ballistic approach.
In the kinetic approach, one includes the interactions between particles by solving the Boltzmann equation for the plasma particles
\begin{equation}
    \frac{d f_i}{dt} = -\mathcal{C}[f_i] \, ,
\end{equation}
where $f_i$ are the distributions of the individual particles and $\mathcal{C}$ are the collision terms.
 Then the EoM for the background field~\cite{Moore:1995ua,Moore:1995si, Ai:2025bjw} takes the form
 \begin{align}
 \label{eq:eomGeneral}
  \Box \phi + \frac{\d V_{0}(\phi)}{\d\phi}+\sum_{i}\frac{\d m^2_i(\phi)}{\d\phi}\int_{\vec k, i}\,{f_i(\vec k,x)} + \underbrace{\Delta m^2_{\Pi_\varphi}(x)\phi- (\partial_\mu\phi) \lim_{q^\mu\rightarrow 0}\frac{{\rm Im}{\Pi}^{\rm R}_\phi(q,x)}{q_\mu} }_{\text{production of particles}} =0 \, ,
 \end{align}
where $\Box=\partial_\mu \partial^\mu$,  $f_i(p,x)$ are the particle distribution functions, and $E_i=\sqrt{|\vec{p}_i|^2+m_i^2}$ the particle energies. Here $x$ and $p$ denote general four-dimensional position and momentum coordinates. In this context, $V_0$ denotes the loop-resummed \emph{zero-temperature} potential. The terms in the underbrace comes from the production of particles studied in the \cite{Ai:2025bjw}. 
Recently this process has been partially automatized in the public code WallGo\cite{Ekstedt:2024fyq,vandeVis:2025plm}.

The \emph{total} pressure is given by 
\begin{align}
\label{eq:total_pres}
\mathcal{P} &= \int dr \phi' \bigg(\frac{\d V_{0}(\phi)}{\d\phi}+\sum_{i}\frac{\d m^2_i(\phi)}{\d\phi}\int_{\vec k, i}\,{f_i(\vec k,x)} + \Delta m^2_{\Pi_\varphi}(x)\phi- (\partial_\mu\phi) \lim_{q^\mu\rightarrow 0}\frac{{\rm Im}{\Pi}^{\rm R}_\phi(q,x)}{q_\mu} \bigg)
\nn
    &= \underbrace{\int dr \phi' \bigg(\frac{\d V_{0}(\phi)}{\d\phi}+\sum_{i}\frac{\d m^2_i(\phi)}{\d\phi}\int_{\vec k, i}\,{f^{\rm eq}_i(\vec k,x, T)} \bigg) }_{\text{LTE effects}}
    \nn
    &
    + \underbrace{\int dr \phi' \bigg(\sum_{i}\frac{\d m^2_i(\phi)}{\d\phi}\int_{\vec k, i}\,{\delta f^{}_i(\vec k,x, T)} + \Delta m^2_{\Pi_\varphi}(x, T)\phi- (\partial_\mu\phi) \lim_{q^\mu\rightarrow 0}\frac{{\rm Im}{\Pi}^{\rm R}_\phi(\vec{q},x, T)}{q_\mu} \bigg)}_{\text{out-of-equilibrium effects}} \, ,
\end{align}
where the first line is the $\mathcal{P}_{\rm LTE} $ which was considered in the previous subsection and the second line are the new out-of-equilibrium effects. We will not discuss them further in this paper and only focus on the LTE part, which constitutes in itself a complicated problem. 

We however emphasized that Eq.~\eqref{eq:vacuumSolution} assumes a constant pressure in time and consequently neglects by construction the plasma backreaction. 
These two approaches (ballistic and kinetic) have been recently reconciled in \cite{Ai:2025bjw}.

\section{Review of the LTE steady state approach}
\label{sec:LTE}

\subsection{General framework}
In the previous section, we studied the expansion of a bubble in a fixed 
medium, whose impact on the bubble was captured through thermal corrections 
to the potential. However, the bubble also disturbs the surrounding 
plasma, generating waves within it. This \emph{deformation} of the plasma 
in turn affects the bubble wall expansion itself, giving rise to the 
so-called \emph{hydrodynamic obstruction}. We now turn to the hydrodynamic 
study of the plasma and the resulting pressure on the wall.

The plasma of the early universe, specifically before the QCD phase transition, can be described by a perfect fluid. Indeed, the viscosity, which will appear in the first-order hydrodynamics of the QGP fluid, is expected to be of order $\eta \sim 0.1 s$\cite{Shuryak:2004cy}, where $s$ is the entropy density. This viscosity is extremely small and consequently, the first-order corrections to the perfect fluid are expected to be very suppressed. 

In order to study the evolution of the temperature profile, it is sufficient to consider conservation of energy and momentum (EM).
The full EM tensor is given by 
\begin{align}
    T^{\mu\nu}=(e+p)u^\mu u^\nu- g^{\mu\nu}p \; , \qquad \qquad \partial_\mu T^{\mu\nu} =0
\end{align}
where $p=-V_{\rm eff}$ and $e=V_{\rm eff}-T\partial V_{\rm eff}/\partial T$.
For an $\mathcal{O}(d)$-symmetric solution, these can be written as
\begin{subequations}\label{eq:EMTConservation}
\begin{align}
\label{eq:energyConservation}
    0 &= \partial_t T\frac{\partial V_{\rm eff}}{\partial T} + \partial_t(w\gamma^2 ) + \partial_r(w\gamma^2 v)+\frac{d-1}{r}w\gamma^2 v,\\
    0 &= -\partial_r T\frac{\partial V_{\rm eff}}{\partial T} + \partial_t(w\gamma^2v) + \partial_r(w\gamma^2 v^2)+\frac{d-1}{r}w\gamma^2 v^2,
\end{align}
\end{subequations}
where $v$ denotes the plasma velocity, $\gamma=1/\sqrt{1-v^2}$ is the associated Lorentz factor, and $w=-T\,\partial V_{\rm eff}/\partial T=e+p$ is the enthalpy density. To get a complete description of the system, these plasma equations need to be supplemented by the bubble equation of motion Eq.~\eqref{eq:wallEOM}. 

\subsection{Review of the hydrodynamical obstruction from the static fluid profiles}
\label{sec:staticLTE}

Using these basic equations, we now present the theory of the wall velocity using \emph{local thermal equilibrium} and highlight the most salient results.

\paragraph{Matching across the wall}

The matching conditions follow from the $\nu=0$ and $\nu=3$ components (assuming the wall being along the 3th component) of $\partial_\mu T^{\mu\nu}=0$. Assuming vanishing first time-derivatives of the hydrodynamic variables, we can integrate Eq.~\eqref{eq:EMTConservation} with $d=1$ across the wall to obtain\cite{Ai:2024shx}
\begin{subequations}
\label{eq:junctionAB}
\begin{align}
      w_+\gamma_+^2v_+ &=w_-\gamma_-^2v_-,\label{eq:conditionA}\\
      w_-\gamma_-^2v_-^2+p_-&=w_+\gamma_+^2v_+^2+p_+\,.\label{eq:conditionB}
\end{align}
\end{subequations}

These matching conditions do not allow to solve for the four unknowns $v_\pm$ and $T_\pm$. To close the system we specify an equation of state (EoS). In the \emph{general} Bag model, one allows for (possibly different) vacuum energies $\epsilon_\pm$ in the two phases,
\begin{align}
\label{eq:bag_eos}
    &e_+(T)=a_+  T^4+\epsilon_+,\qquad p_+(T)=\tfrac{1}{3}a_+ T^4-\epsilon_+,\nonumber\\
    &e_-(T)=a_-  T^4+\epsilon_- ,\qquad\;\  p_-(T)=\tfrac{1}{3}a_- T^4-\epsilon_-, 
\end{align}
 where $\epsilon_\pm$ is the vacuum (zero-temperature) potential energy densities of the two phases and where a subscript ``$\pm$'' denotes quantities just in front of ($+$) and just behind ($-$) the wall. We also defined $a$, which encodes the number of \emph{relativistic} degrees of freedom in each phase, 
$
    a_{\pm} = \frac{\pi^2}{30} \sum_{\text{light } i} 
\left[ N_i^b + \frac{7}{8} N_i^f \right] \, ,
$
where $N^{b(f)}_i$ are the number of light bosons (fermions) dofs.  
 Eliminating $p$ and $w$ in Eq.~\eqref{eq:junctionAB} and using the bag EoS, yields the standard relation between the fluid velocities $v_\pm$ in the wall rest frame,
\begin{equation}
\label{eq:vplus_vminus_relation}
v_+(v_-, \alpha_+) = \frac{1}{1+\alpha_+} \bigg[\bigg(\frac{v_-}{2}+ \frac{1}{6v_-}\bigg) \pm \sqrt{\bigg(\frac{v_-}{2}+ \frac{1}{6v_-}\bigg)^2 + \alpha_+^2 +\frac{2}{3}\alpha_+- \frac{1}{3}}\bigg] .
\end{equation}
The $-\ (+)$ correspond to the branches of deflagrations (detonations) respectively, and we defined 
\begin{equation}
\bar{\theta} = \left( e - \frac{1}{c_{s,-}^2}p \right),
\qquad
\alpha_{\bar\theta}(T) = \frac{\Delta \bar{\theta}(T)}{3 \, \omega_+(T)} \, \, , 
\end{equation}
and
\begin{equation}\label{eq:defAlpha}
    \alpha_+ \equiv \alpha_{\bar\theta}(T_+), \qquad \alpha_n \equiv \alpha_{\bar\theta}(T_n), \qquad  \alpha_{p} \equiv \frac{\nu\Delta V_{\rm eff}(T_n)}{3w_+ (T_n)}   \, ,
\end{equation}
with $\nu=1+1/c_{s,-}^2$.
The two definitions $\alpha_n$ and $\alpha_p$ are related by the exact relation
\begin{equation}\label{eq:defPsi}
     \alpha_n = \alpha_{p} + \frac{1-\Psi}{3}\, ,\qquad \Psi(T)=\frac{w_-(T)}{w_+(T)}\, ,
\end{equation}
and can therefore be used interchangeably. Some situations, however, are described more transparently by one definition than the other. For example, the hydrodynamic equations are sourced by the latent heat, which is approximately $\bar\theta$, making $\alpha_n$ the more appropriate measure of the wave amplitude. The force driving the wall forward, by contrast, is $\Delta V_{\rm eff}$, making $\alpha_p$ better suited to describing the wall velocity.

\begin{figure}[t]
 \centering  \includegraphics[width=0.48\textwidth]{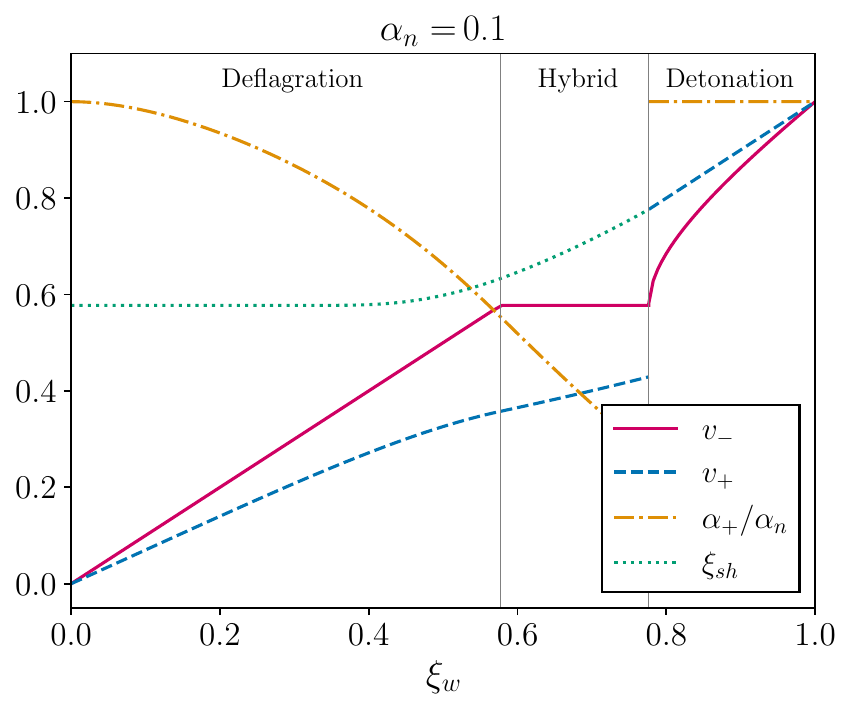}\ \includegraphics[width=0.488\textwidth]{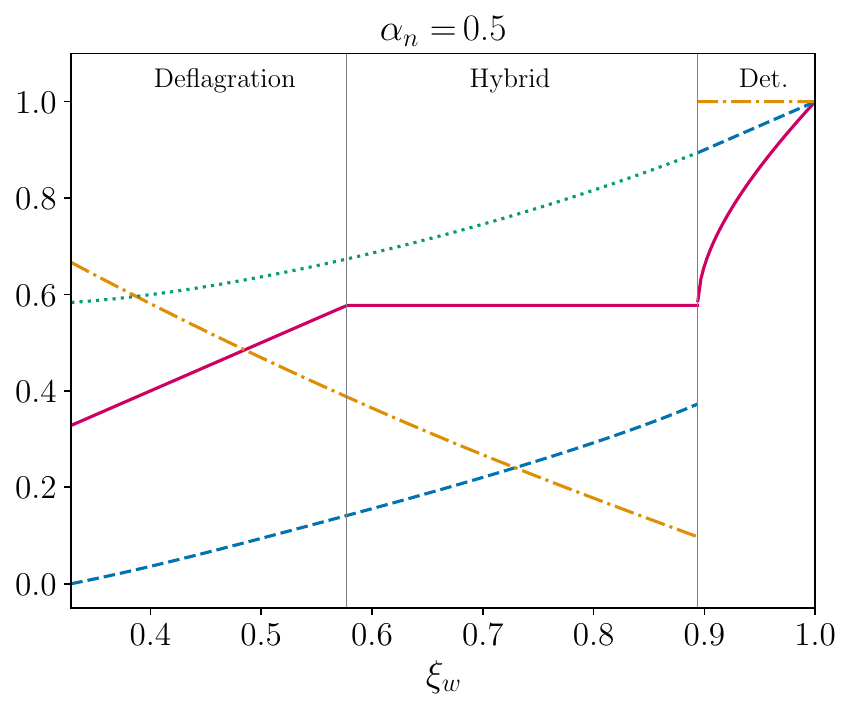}\\ 
 \caption{$v_-, v_+, \alpha_{\theta,+}$ and the shock velocity $\xi_{\rm sh}$ as a function of the wall velocity. The plot is separated in three regimes: detonations, for wall velocity larger than the Jouguet velocity, deflagration, for wall velocity smaller than the speed of sound, and hybrid, for the regime in between. }
 \label{fig:quantities direct}
 \end{figure}

Another type of hydrodynamical discontinuity is the shock propagating in front of the wall, across which the plasma does not change nature and which correspond to the case $\alpha_+ =0$. Consequently, we obtain $v_{\rm sh,+} v_{\rm sh,-} =1/3$. This shock lies ahead of the wall position in the case of the hybrid and deflagration solutions. The steady state profiles are static in time (once they are expressed in the self-similar coordinate) and thus can be reexpressed in the form of a single \emph{self-similar} variable $\xi\equiv r/t$. In Fig.~\ref{fig:quantities direct}, we present the velocity of the fluid $v_+, v_-$ and the position of the shock $\xi_{\rm sh}$ as a function of the wall velocity $\xi_w$ for two examples.

Cosmological phase transitions have three different hydrodynamical regimes: detonations, for wall velocity larger than the Jouguet velocity $v_J$, deflagration, for wall velocity smaller than the speed of sound, and hybrid, for the regime in between. The Jouguet velocity is defined as the wall velocity catching up the shock wave
\begin{align}\label{eq:jouguet}
    v_J=\frac{1}{\sqrt{3}}\left(\frac{1+\sqrt{\alpha_{n}(2+3\alpha_{n})}}{1+\alpha_{n}}\right) \, .
\end{align}

\paragraph{Imposing the conservation of entropy in the wall}

In the limit in which one can assume that the plasma inside the bubble wall remains close enough to equilibrium~\cite{Balaji:2020yrx, Ai:2021kak, Ai:2023see,Ai:2024shx}, a further matching condition allows to solve for the wall velocity. The evolution of thermodynamical quantities inside the wall is thus well captured by a hydrodynamical \emph{continuous} wave. As it is well known, continuous waves conserve the entropy current $s u^\mu$, such that $
\partial_\mu(s u^\mu) = 0 
$, 
\begin{equation} 
\label{eq:LTE-matching_1}
   \partial_\mu (s u^\mu) = 0\quad\Rightarrow\quad  s_+ \gamma_+ v_+ =s_- \gamma_- v_- \quad \text{(matching condition from entropy current conservation)}.
\end{equation}
and, using Eqs.~\eqref{eq:junctionAB}, we can recast the ratio of temperatures in the ratio of velocities
\begin{align}
 \label{eq:LTE-matching}
\frac{T_+}{T_-}=\frac{\gamma_-}{\gamma_+}\,.
\end{align}
 This new matching condition fixes the last parameter left unconstrained by hydrodynamics, the wall velocity. 
Finally, the complete pressure on the bubble wall in the LTE regime  is given by\cite{Ai:2021kak, Ai:2023see, Ai:2024shx}
\begin{equation}
\label{eq:pres}
\mathcal{P}_{\rm LTE}  = \frac{w_+}{4} \bigg(4 \gamma_+^2 v_+(v_+-v_-) + 1 - \Psi\bigg(\frac{\gamma_+}{\gamma_-} \bigg)^4 - 3 \alpha_+(\xi_w)\bigg) \, . \qquad \text{(LTE pressure)}
\end{equation}

With this equation at hand, we can study the pressure for the direct and the inverse phase transitions.
First of all, we can study this relation in the case of a just-nucleated bubble. The bubble is taken to be nucleated at rest with a vanishing wall velocity, $\xi_w \to 0$, and correspondingly $v_+ = v_- \to 0$, consequently we have that  
\begin{equation}
\mathcal{P}_{\rm LTE}(v_+)\bigg|_{\xi_w = v_+ = v_- \to 0} \to \frac{w_+}{4} \bigg( 1 - \Psi - 3 \alpha_n\bigg) \, . 
\end{equation}
If this pressure is positive, then the bubble cannot expand from the start. This can be understood from the fact that in the region of the parameter space with 
\begin{equation}
\label{eq:cond_exp}
\alpha_n < \frac{1-\Psi}{3} \quad \Rightarrow \quad \alpha_p < 0 \qquad \qquad  \text{(Nucleation thermodynamically disfavored)} \, ,
\end{equation}
the bubble nucleation itself is thermodynamically disfavored~\cite{Espinosa:2010hh}.

\begin{figure}[t]
 \centering  \includegraphics[width=0.45\textwidth]{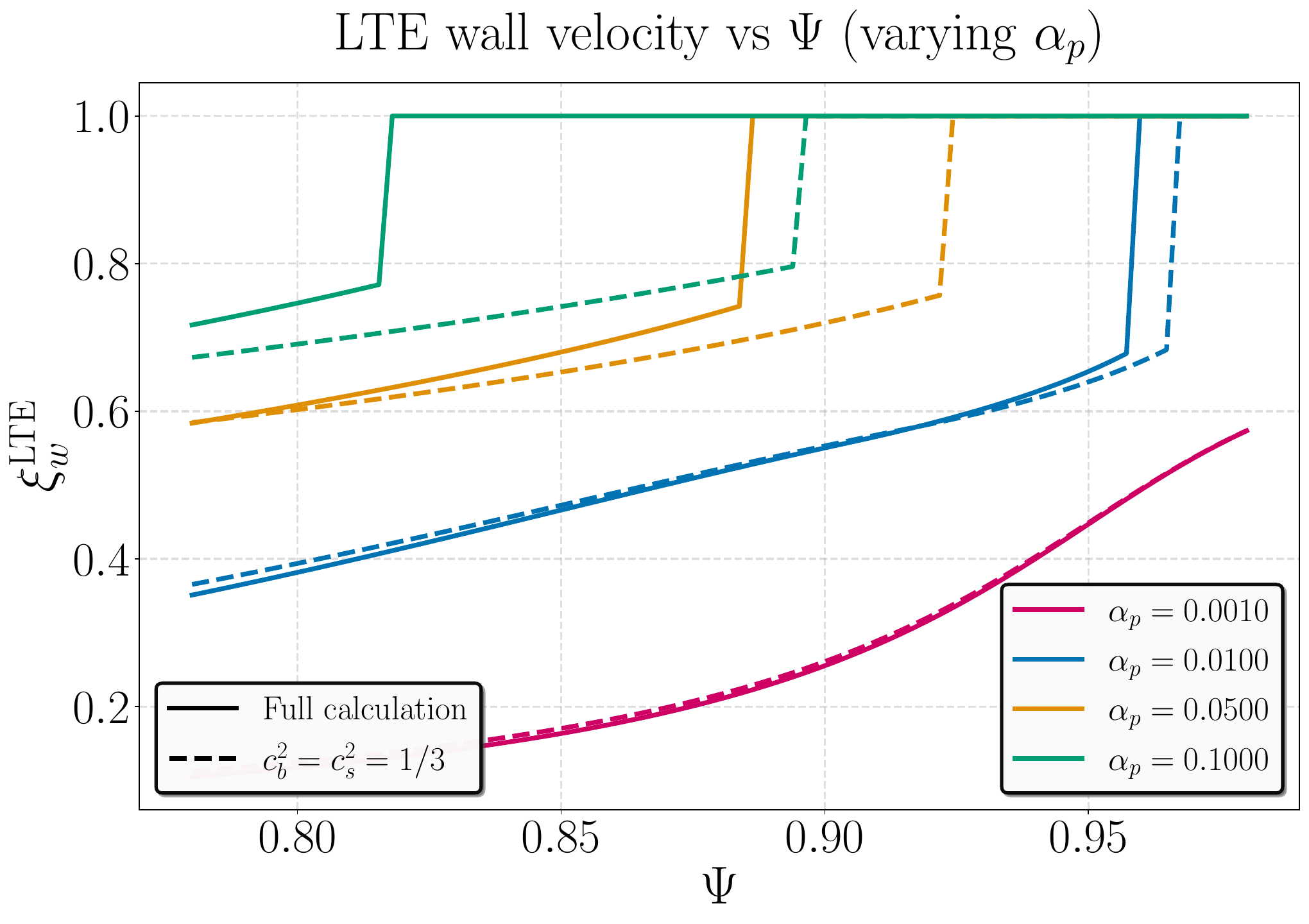}\
 \includegraphics[width=0.45\textwidth]{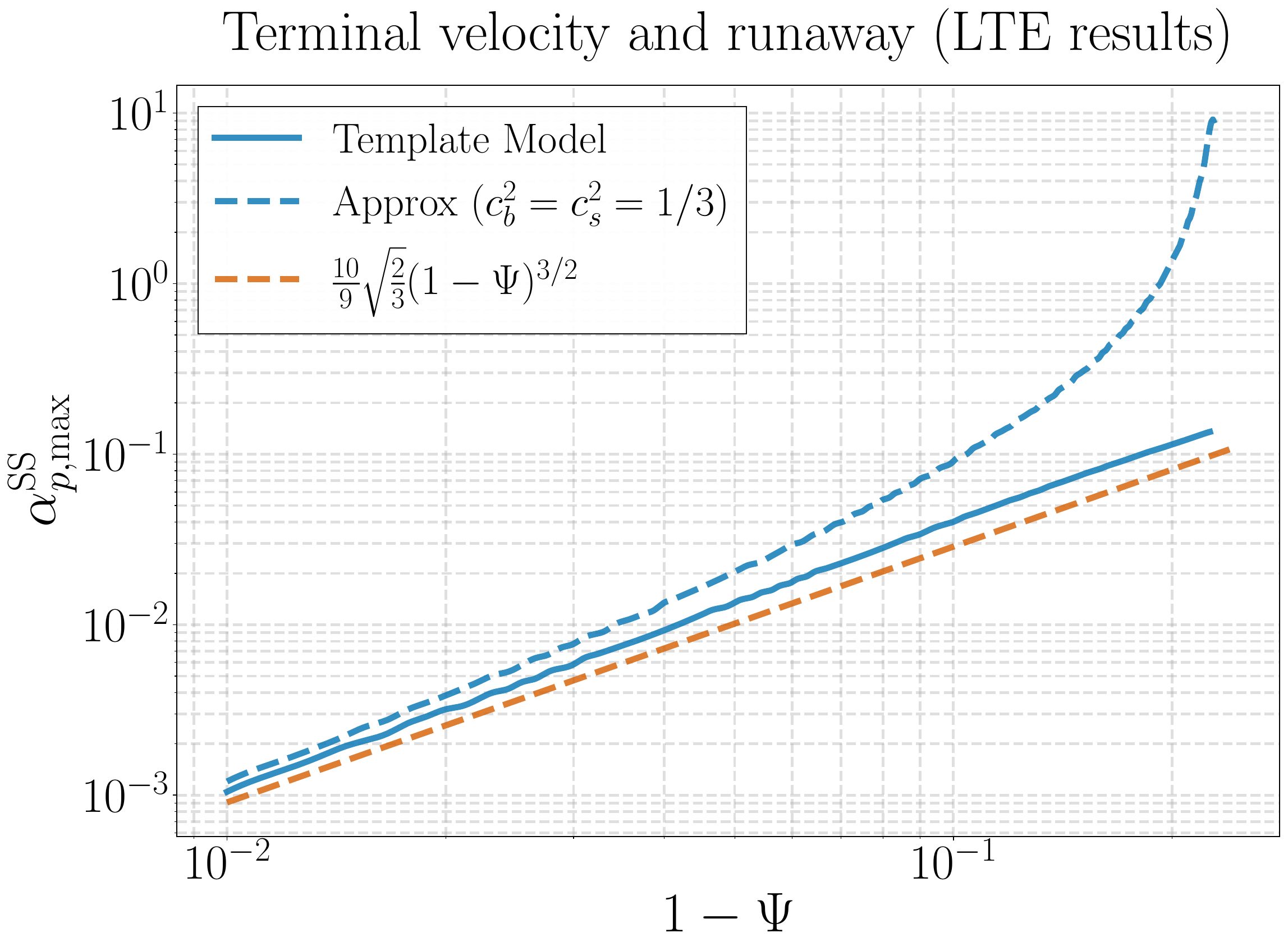}\
 \caption{\textbf{Left panel}: Wall velocity $\xi_w^{\mathrm{LTE}}$ versus phase transition strength $\Psi$ for varying latent heat parameters $\alpha_p$. Solid lines represent full calculations with temperature-dependent sound speed parameters; dashed lines employ the simplifying assumption $c_{s,\pm}^2 = 1/3$ while for solid lines we draw the speed velocity directly from the potential (\ref{eq:bagPotential}). Results span $\alpha_p \in \{0.001, 0.01, 0.05, 0.1\}$ with each color denoting a distinct value. \textbf{Right panel}: Boundary between a terminal velocity (below the curve) regime and a runaway regime (Above the curve) within the LTE formalism for the approximate case of $c_{s,\pm}^2 = 1/3$ (dashed blue) and the case where the speed of sound is derived from the potential (\ref{eq:bagPotential}) (solid blue). The region above the curve implies runaway, while the region below implies terminal velocity regime. }
 \label{fig:LTE_velo}
 \end{figure}

It was shown that the pressure peaks at the Jouguet velocity\cite{Ai:2023see, Ai:2021kak, Ai:2024shx}, reflecting the fact that the heating of the wall is maximal at the Jouguet velocity. This implies that there are either no solution to the equation $\mathcal{P}_{\rm LTE} = 0$ or two.  The first case leads to a runaway or a terminal velocity controlled by strongly out-of-equilibrium (for example ballistic) effects. The second case, where there are in principle two solutions to the equation, leads to a terminal velocity controlled by thermodynamical effects. The first thing to note is that the second solution (at larger $\xi_w$) is unstable within the LTE approximation. This leaves only the first solution (with lower $\xi_w$) as a stable solution. In principle, also in this case, there might exist a second stable solution at larger wall velocity stabilized by out-of-equilibrium effect.  

We show the velocity of these solutions in Fig.~\ref{fig:LTE_velo}. We always select the solution with the smallest velocity in the plot (the stable solution) and when the velocity jumps to one, it means that the equation $\mathcal{P}_{\rm bubble} = 0$ is never satisfied and the wall runs away. Dashed lines employ the simplifying assumption $c_{s,\pm} = 1/3$ while for solid lines we draw the sound speed directly from the potential (\ref{eq:bagPotential}), which yields $c_{s,+}^2=1/3$ and $c_{s,-}^2=\Psi/(2+\Psi)$. We observe that the two curves start to drift apart at larger $\alpha_p$, reflecting the fact that the bag EoS is not an appropriate approximation for strong PTs.

The presence of the hydrodynamical obstruction leads to a boundary value of $\alpha_p^{\rm max}$ as a  function $\Psi$ such that any value of $\alpha_p < \alpha_p^{\rm max}(\Psi)$ has a solution of $\mathcal{P}_{\rm bubble} = 0$ for $\xi_w < v_J$. This function has been built explicitly in \cite{Ai:2024shx} for the bag EoS and it has been found that it diverges for $\Psi \lesssim 0.77$. We reproduce this calculation in Appendix \ref{app:alphaMax}. This seems to lead to the counterintuitive consequence that, within the bag EoS, detonations are impossible for any value of $\Psi \lesssim 0.77$\cite{Ai:2024shx}.

\subsection{Hydrodynamical simulations and approach to steady states}
\label{sec:hydro}

In the remainder of this section, we present hydrodynamic simulations of the plasma evolving in the presence of a prescribed (fixed shape) scalar field. Specifically, we set
\begin{align}\label{eq:fixedScalar}
    \phi(r,t) = \frac{\phi_0}{2}\bigg[1-\tanh\bigg(\frac{\gamma_{\sss R}(r^2-R^2(t))}{rL}\bigg)\bigg],
\end{align}
and solve the conservation equations~(\ref{eq:EMTConservation}) for $v$ and $T$ using the pseudospectral method described in Appendix~\ref{sec:numericalAlgorithm}. The VEV $\phi_0$, the wall thickness $L$, and the wall position $R(t)$ are fixed by hand to model different situations. Although these simulations do not solve the scalar EoM, they provide useful intuition about the shock formation time scale.

In the purely hydrodynamical picture presented in the former subsections, the pressure was determined using the boundary conditions. We reiterate the definition of the LTE pressure already obtained in Eq.~\eqref{eq:total_pres},
\begin{equation}
 \mathcal{P}_{\rm LTE}\equiv  \int dr\, \partial_r \phi \bigg(  \frac{\d V_0(\phi)}{\d\phi}+\sum_i\frac{\d m^2_i(\phi)}{\d\phi}\int \frac{\d^3{\bf p}}{(2\pi)^32E_i}\,f_i(p,x, T) \bigg).
\end{equation}
However, in the context of plasma which is always in local thermal equilibrium, one can again interpret $f_i(p,x, T) = f_i^{\rm eq}(p,x, T)$ and the two terms in the equation above combine to constitute the fully thermally corrected (and loop-resummed) potential, 
\begin{equation}
\label{eq:Pbubble}
    \mathcal{P}_{\rm LTE}(t) = \int_{0}^{+\infty} dr\, \partial_r \phi(r,t) \, \left.\frac{\partial V_{\rm eff}(\phi,T)}{\partial \phi}\right|_{(\phi(r,t), T(r,t))} \, .
\end{equation}
One can understand this equation as the change in pressure across the wall, being captured by the derivative of the scalar field, which is only nonzero inside the bubble wall. In all that will follow, we will use a toy model potential of the form presented in Appendix \ref{sec:potential} and will perform the substitution $V_{\rm eff} \to V_{\rm HT}$.

Armed with the hydrodynamical equations \eqref{eq:EMTConservation}, we can now turn to  the question ``what could fail with the {\it static fluid profile} determination of the velocity?'' 

\paragraph{The approach to the steady state}
As we already emphasized, hydrodynamics carries an inherent response timescale, which can be understood as the typical timescale over which waves are formed when a spatial gradient of pressure is turned on, or equivalently when a source of heat is turned on. Schematically, the relevant hydrodynamical equation takes the form $\partial_t v \sim (\partial_r p)/p$.

\begin{figure}[t]
 \centering  
 \includegraphics[width=0.488\textwidth]{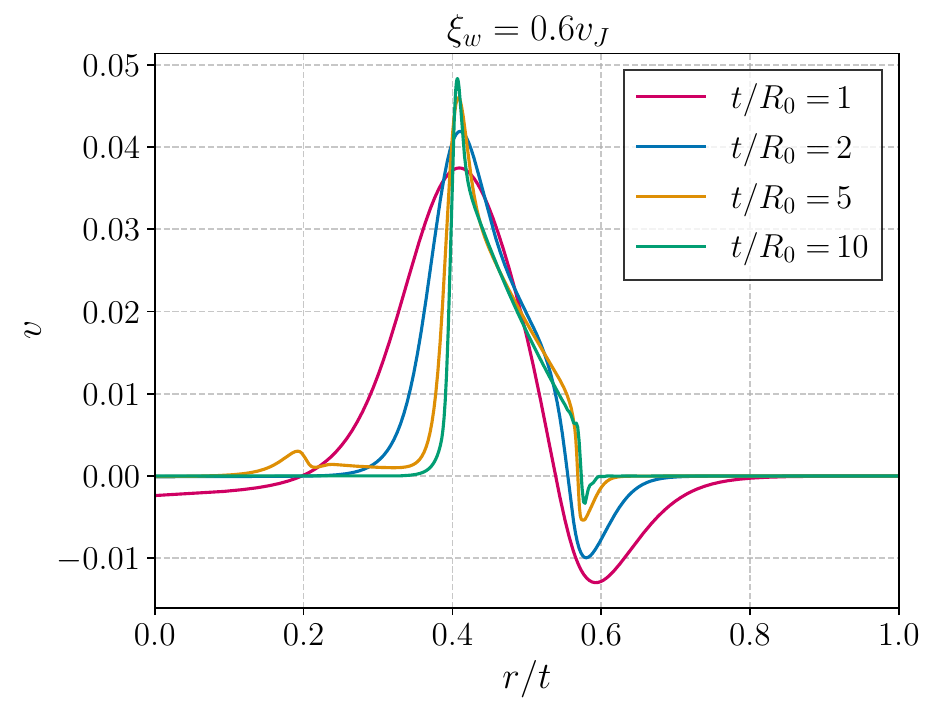}\includegraphics[width=0.48\textwidth]{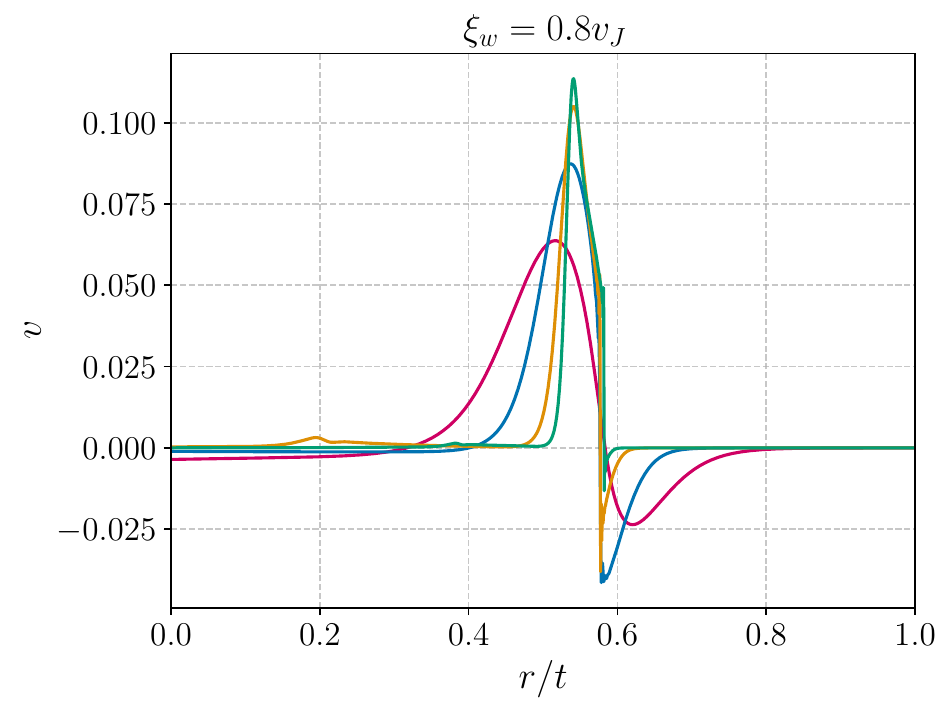}\ \\
\includegraphics[width=0.48\textwidth]{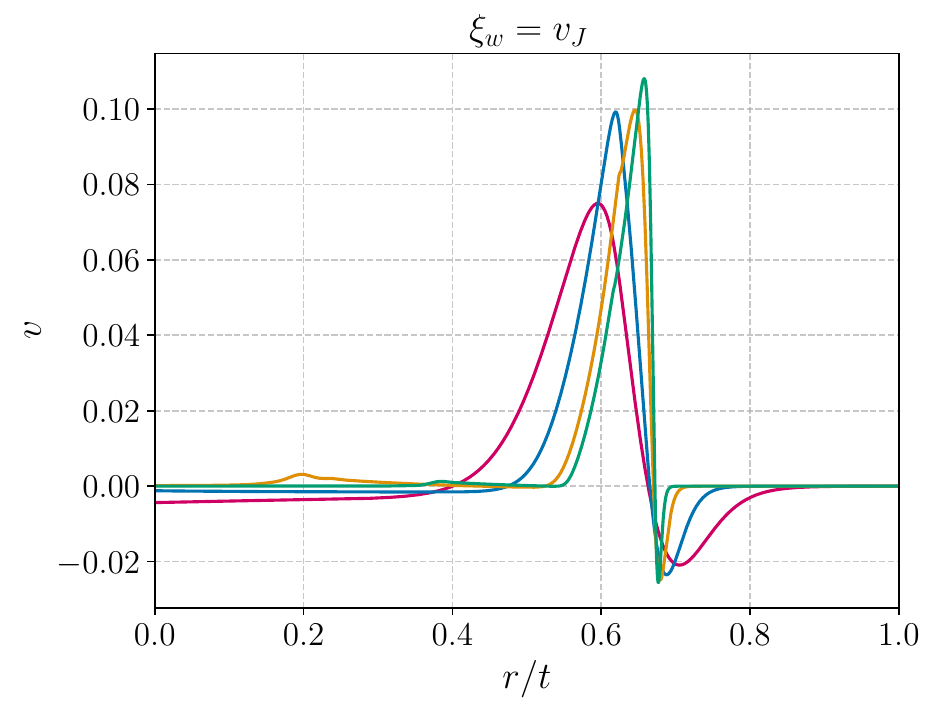}\ \includegraphics[width=0.488\textwidth]{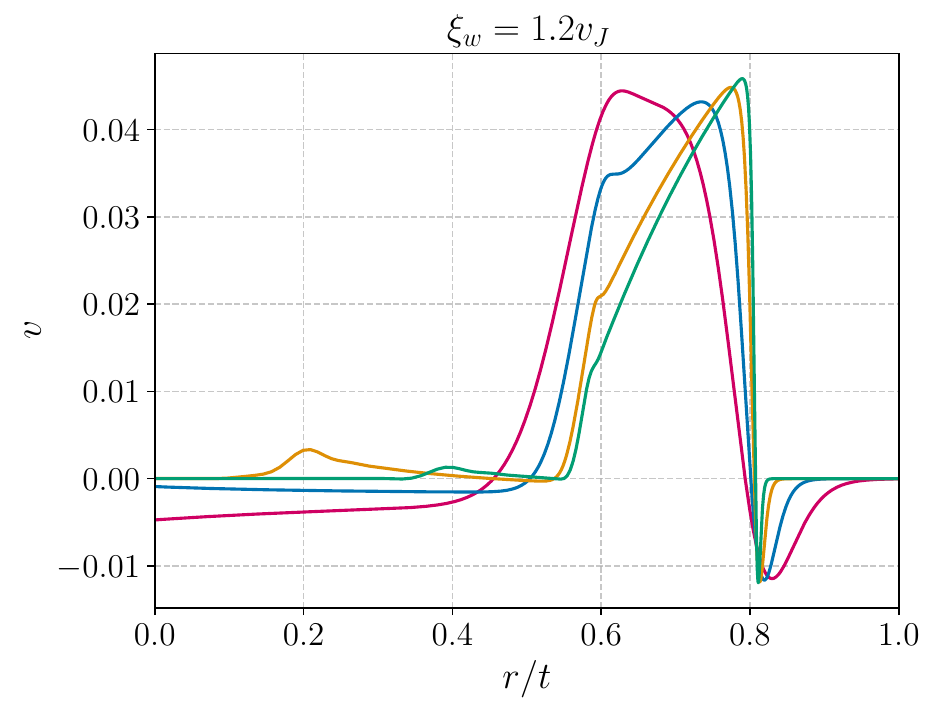}\\
 \caption{Study of the approach to scaling of the for $\alpha_p=0.005$, $\Psi=0.95$, $L=1/T_n$ and velocities $\xi_w = (0.6,0.8, 1, 1.2) \times v_J$, where $v_J$ is the Jouguet velocity corresponding to the transition with a given $\alpha_p$.}
 \label{fig:approach_ss}
 \end{figure}
\begin{figure}[t]
 \centering  
 \includegraphics[width=0.45\textwidth]{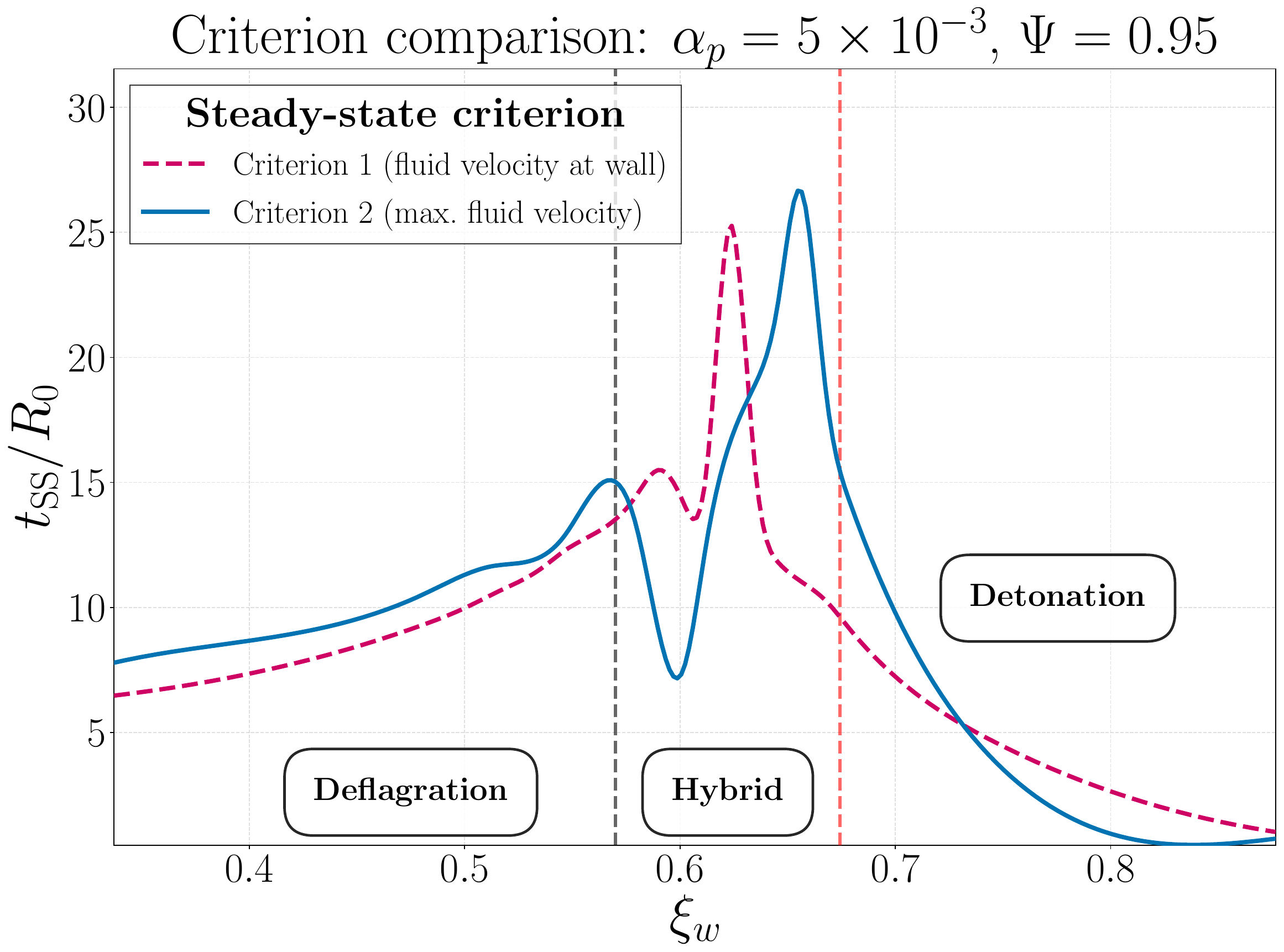}
\includegraphics[width=0.45\textwidth]{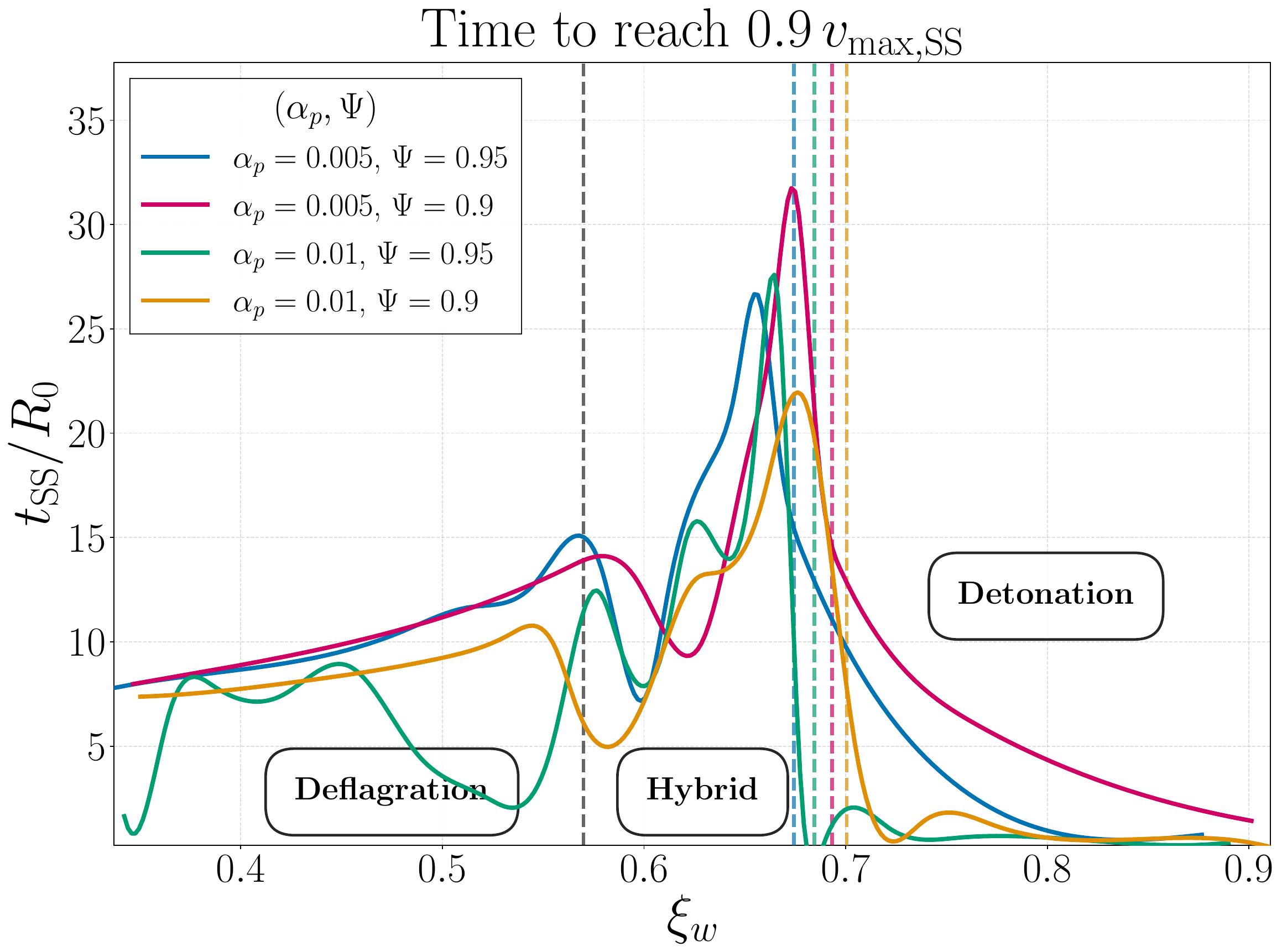}\ \\
 \caption{\textbf{Left panel:} Time to reach the steady state $t_{\rm SS}/R_0$
    as a function of the wall velocity $\xi_w$, comparing Criterion~1
    (fluid velocity at the wall) and Criterion~2 (maximum fluid velocity).
    \textbf{Right panel:} Same quantity using Criterion~2 only, for four
    combinations of the parameters $(\alpha_p, \Psi)$. Vertical dotted
    lines indicate the Jouguet velocity $v_J$ for each curve. We typically observe a peak in the formation time close to the speed of sound. }
 \label{fig:approach_ss_tending}
 \end{figure}

A first question that we might ask is: if we make a planar bubble appear magically and let it expand at constant velocity, how long would it take to reach the steady state profile? We will call this time of formation of the steady state: \emph{the formation time}. To obtain an estimate of the formation time, we proceed in the following way: we use the 1+1 hydrodynamical simulation presented in the former subsection and fix a constant velocity $\xi_w$ with the choice $R(t)=R_0+\xi_w t$. We then study the approach to steady state of the fluid profiles, which is shown in Fig.~\ref{fig:approach_ss}.

In Fig.~\ref{fig:approach_ss_tending},
to quantify the time required for the system to reach a steady state,
we monitor the temporal evolution of the fluid profile for each wall
velocity $\xi_w$ and extract the relaxation time $t_{\rm SS}$ using two
complementary criteria. Criterion~1 tracks the convergence of the fluid velocity at the position of the wall, while Criterion~2 monitors the stabilization of the maximum
fluid velocity in the profile; the steady state is declared reached once
the relative variation of the relevant quantity falls below a fixed
threshold, chosen to be 90\% of the asymptotic solution. As shown in Fig.~\ref{fig:approach_ss_tending} (left), the two criteria
yield moderately consistent estimates of $t_{\rm SS}$, confirming the robustness of
our definition (notice however the offset in the curves). We therefore adopt Criterion~2 throughout the remainder
of the analysis and, in Fig.~\ref{fig:approach_ss_tending} (right), explore its
dependence on the model parameters by scanning four representative
combinations of the strength parameter $\alpha_p$ and the ratio $\Psi$. 

We observe that the required time ranges between $t/R_0 \in [2, 25]$, depending on the imposed velocity $\xi_w$. One important outcome is that the convergence is the fastest for the supersonic detonation and deflagration, while it is by far the slowest for hybrids with velocity close to the Jouguet velocity. We however observe that this trend is quite independent on the explicit values of $\Psi$ and $\alpha_p$.

\paragraph{Pressure for a constant wall velocity}

\begin{figure}[t]
    \centering
    \includegraphics[width=0.49\textwidth]{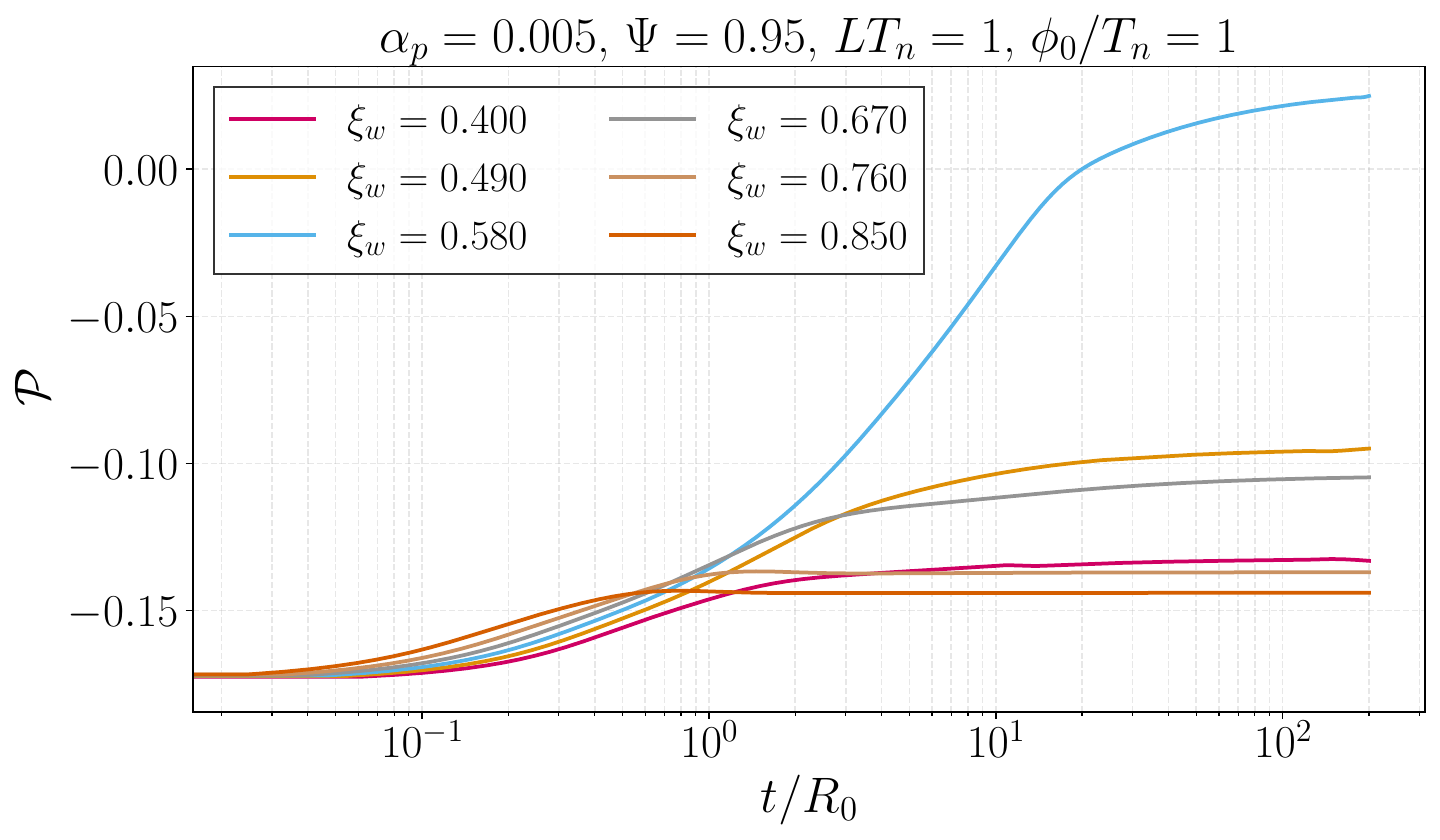}
\includegraphics[width=0.49\textwidth]{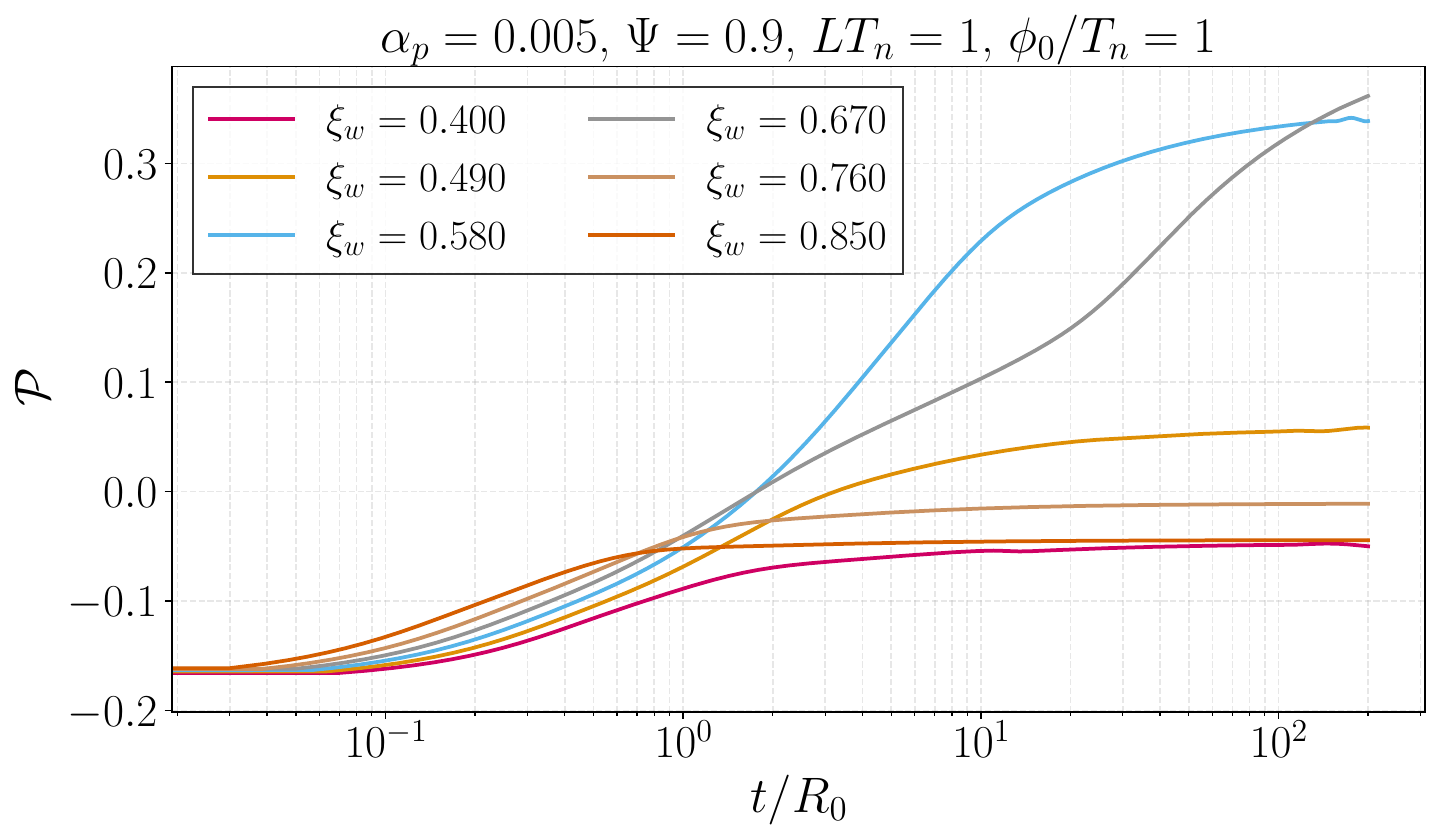}
    \caption{
       Pressure evolution $\mathcal{P}(t)$ as a function of rescaled time $t/R_0$ for different constant wall velocities 
        $\xi_w \in [0.4, 0.85]$ with $\alpha_p = 0.005$, and $\Psi = 0.95$ (Left), $\Psi = 0.9$ (Right). The simulations show rapid transient dynamics at early times, followed by 
        convergence to a quasi-steady state by $t/R_0 \approx 10$. Each color represents a different wall velocity.  
        }
    \label{fig:pressure_vs_time}
\end{figure}

Finally, before closing this subsection, we study how the pressure on the bubble wall, defined in Eq.~(\ref{eq:Pbubble}), converges at late times. Figure~\ref{fig:pressure_vs_time} displays the time evolution of the pressure $\mathcal{P}_{\rm LTE}(t)$ 
for six different wall velocities. Several key observations emerge:
at early times ($t/R_0 \lesssim 10$), the pressure increases quickly for all the velocities. 
 After this transient phase, the pressure converges to an approximately constant value, indicating that a quasi-steady-state regime has been reached. We observe two different regimes of convergence: a fast regime and a slow regime.
 For velocities $\xi_w < 0.8 v_J$ or $\xi_w > v_J$, the asymptotic late time value of $\mathcal{P}^{}_{\rm LTE} (t)$, which we call $\mathcal{P}^{\rm SS}_{\rm LTE}$, SS standing for ``steady state'', is reached within $t/R_0\sim 2-4$, which implies very fast convergence of the pressure. However, for velocities $v_J>\xi_w > 0.8 v_J$, the convergence to SS  is reached within $t/R_0\sim 10-30$, which is a much slower regime.  
 
 This result is consistent with the conclusion reached in Fig.~\ref{fig:approach_ss} and Fig.~\ref{fig:approach_ss_tending}, where it was found that a profile very close to the steady state fluid profile was reached after around $t/R_0\sim 20-30$ for hybrids and after $t/R_0\sim 2-10$ for detonations and deflagrations.

\subsection{Pressure evolution and wall velocity}

In subsection \ref{sec:staticLTE}, we described the pressure function $\mathcal{P}_{\rm LTE}(\xi_w)$ assuming that the fluid profile was exactly tracking the steady state at all times during the bubble acceleration. The pressure profile $\mathcal{P}_{\rm LTE}(\xi_w)$ was then shown to display a maximum at a velocity very close to the Jouguet velocity, see for example the reference\cite{Ai:2023see} for further examples. In the last subsection \ref{sec:hydro}, we however argued on intuitive grounds that it was unlikely that the fluid profile could efficiently track the steady state profile, as the timescale of acceleration (the timescale over which the velocity changes) competes, or is even shorter, that the timescale of the wave formation. 

In this subsection, we turn to a first qualitative assessment of this claim. We will run purely hydrodynamical simulations with an accelerating fixed Higgs wall described by Eq.~\eqref{eq:fixedScalar}. As they do not include the backreaction of the plasma on the wall, the simulations we are running in this section should not be taken as quantitatively exact. We will see that they nevertheless provide a useful intuition on the problem at hand. 

We impose a constant proper acceleration on the wall, with the wall position given by 
\begin{equation}
    R(t) = R_0+\frac{\sqrt{1+A^2 t^2} -1}{A} \, ,
\end{equation}
 where $A$ is a given acceleration, which we will express as a fraction of the physical initial acceleration $A_0 \equiv \ddot R(0) = \frac{9 L \alpha_p w_n}{4 \phi_{0}^2}$  given by $\ddot{R}(t=0)$ in Eq.~\eqref{eq:wallEOM1}. We will vary by hand the ratio $A/A_0$ to parameterize different physical situations of the early wall acceleration. Here, $L$ is the thickness of the wall, $w_n$ is the enthalpy density at the nucleation temperature, and $\phi_0$ is the vacuum expectation value. The wall velocity as a
function of time is given by the relativistic expression
\begin{equation}
    \xi_w(t) = \frac{A\,t}{\sqrt{1 + A^2 t^2}} \, ,
    \label{eq:vw_acc}
\end{equation}
which asymptotically approaches $\xi_w \to 1$ as $t\to\infty$,
while remaining subluminal at all finite times. 
Conversely, the time necessary to reach the Jouguet velocity is given by 
\begin{align}\label{eq:jouguetTime}
   \xi_w(t_J) = v_J \quad \rightarrow \quad  t_J &= \frac{\left(1+\sqrt{\alpha_n(2+3\alpha_n)}\right)}
{A\left[3(1+\alpha_n)^2-\left(1+\sqrt{\alpha_n(2+3\alpha_n)}\right)^2\right]^{1/2}}
\\ \nonumber
&\approx \frac{1}{\sqrt{2}A}\left(1 + \frac{3}{\sqrt{2}}\,\alpha_n^{1/2} 
+ \mathcal{O}(\alpha_n)\right) \, . 
\end{align}

\begin{figure}[t]
 \centering  \includegraphics[width=0.47\textwidth]{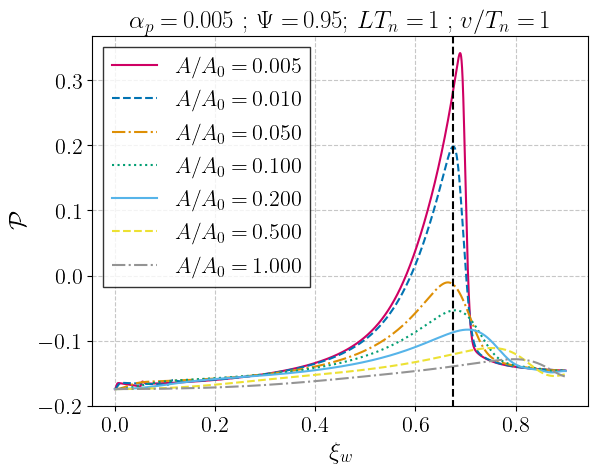}\ \includegraphics[width=0.52\textwidth]{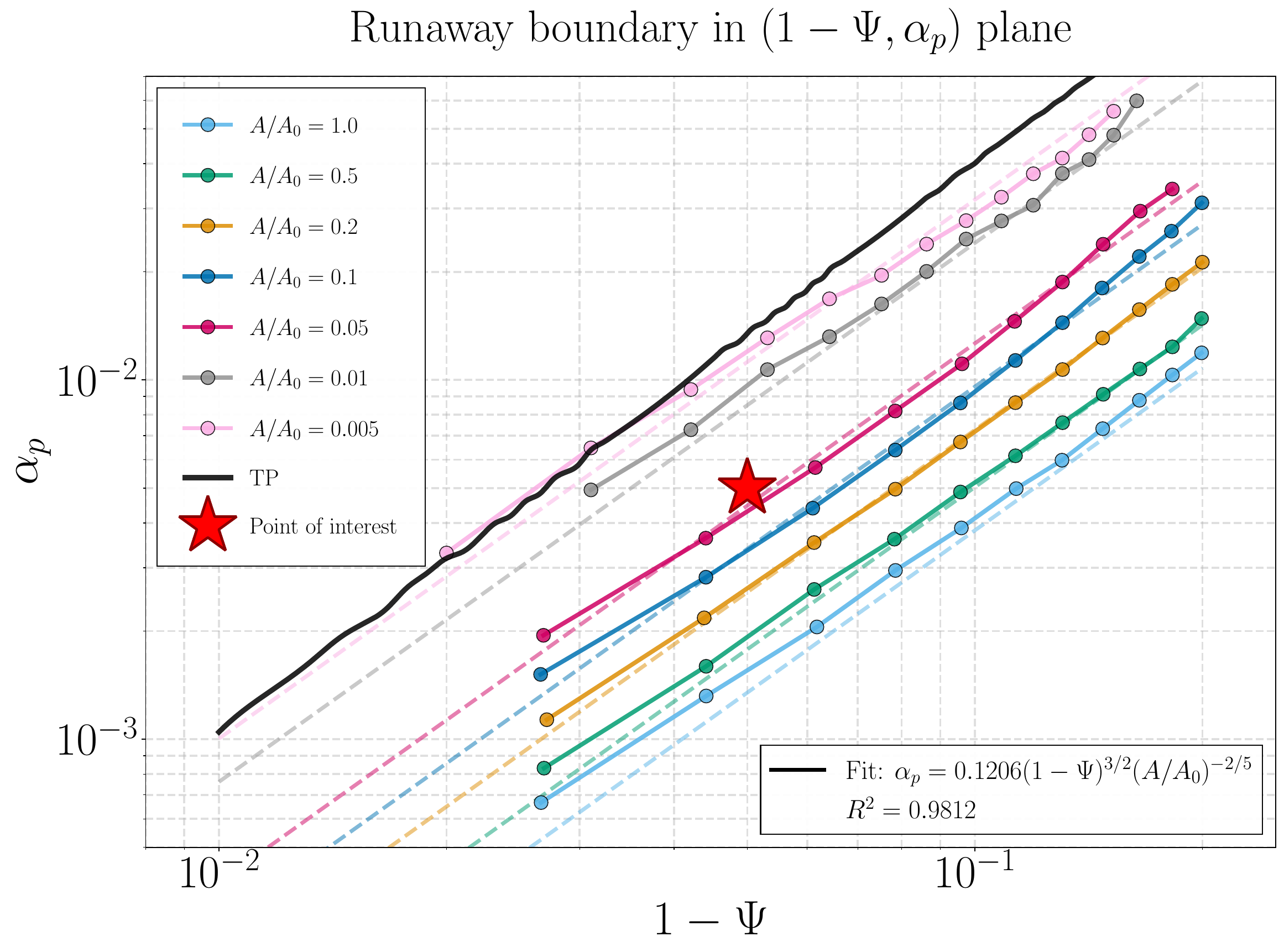}\
 \caption{\textbf{Left panel}: On the left panel we show the evolution of the pressure on the bubble using Eq.~\eqref{eq:pres} for different accelerations. The vertical dashed line shows the Jouguet velocity. \textbf{Right panel}: We show the boundary in the plane $(\Psi, \alpha_p)$ between the runaway (above the contour line) and the terminal velocity (below the contour line) regime for different acceleration histories. We compare this boundary with the SS boundaries obtained in the template model (TP) in Fig.~\ref{fig:LTE_velo}. We also obtain the best fit for the boundary in the form: $\alpha_p^{\rm boundary} \simeq 0.12\,(1-\Psi)^{3/2}\,\left(A/A_0\right)^{-2/5}$. The dashed lines are drawn from the fit presented in the lower right legend. The red star represent the value of the point studied on the left panel. }
 \label{fig:boundary}
 \end{figure}

On the left panel of Fig.~\ref{fig:boundary}, we present the pressure evolution for different acceleration histories. As one can observe, the pressure (namely at the Jouguet velocity) reaches larger and larger values when the acceleration, controlled by the ratio $A/A_0$ is milder. This can be understood intuitively by remembering that slower acceleration gives more time for the shock wave to form, allowing it to generate more pressure on the wall. This is an explicit consequence of Eq.~\eqref{eq:jouguetTime}.

The right panel of Fig.~\ref{fig:boundary} displays the runaway boundary in the $(\Psi, \alpha_p)$ plane
for several values of the wall acceleration parameter $A/A_0$.
We determine the boundary between the \emph{runaway} and
\emph{terminal velocity} regimes of a bubble wall in the
$(\Psi,\,\alpha_p)$ plane, with $\alpha_p$ and $\Psi$ defined in Eqs.~\eqref{eq:defAlpha} and \eqref{eq:defPsi}.  A wall is classified as
\textbf{runaway} if $\max_t\mathcal{P}(t)\leq 0$, and as
\textbf{terminal} otherwise.
For fixed $A/A_0$, the critical phase-transition strength $\alpha_p$ decreases monotonically
as $\Psi \to 1$.
Increasing $A/A_0$ shifts the boundary downward, meaning that a stronger driving acceleration
lowers the threshold $\alpha_p$ needed for runaway.
A global power-law fit across all curves yields
\begin{equation}
    \alpha_p^{\rm boundary} \simeq 0.12\,(1-\Psi)^{3/2}\,\left(\frac{A}{A_0}\right)^{-2/5},
    \label{eq:fit}
\end{equation}
which reproduces the numerical data with $R^2 \approx 0.98$, confirming the separable power-law
dependence on both $\Psi$ and $A/A_0$ over the entire parameter range explored. Of course, as $A$ is decreased low enough, we expect it to saturate to the value of the SS analysis. Typically, the fit starts to deviate at very small $1-\Psi \lesssim 4\times 10^{-2}$ value from the trend presented above (see Eq.~\eqref{eq:fit}). The acceleration $A/A_0 \sim 0.005$ is slow enough for the boundary to tend to the SS boundary within the template model. 

Before to move to more realistic dynamical simulations, we emphasize that we do not expect quantitative matching between the fixed scalar and the dynamical scalar cases, but merely qualitative agreement.

\subsection{Conclusion on the static LTE approach}

As we have discussed in this section, even in systems presenting a steady state attractor solution, as for the expansion of bubbles with constant velocity, there is an intrinsic timescale associated to the formation of this profile. 

This timescale can however be of the same order, or even longer than the timescale associated to the bubble wall acceleration in Eq.~\eqref{eq:eqoM} where we saw that in the absence of external pressure $\gamma \propto R/R_0$. We can conclude from these considerations that it is very dubious that the ``steady state tracking'' hypothesis is generically fulfilled. In fact, we already observed within the framework of non-dynamical accelerating wall simulations, that the boundary between the terminal velocity and the runaway moves away from the SS case as the wall accelerates faster and faster (see Fig.~\ref{fig:boundary} and the associated fit of the boundary Eq.~\eqref{eq:fit}).

We now aim to systematize this intuition by developing a dynamical scalar+fluid coupled 1+1 simulation to understand the dynamics of the formation of the fluid profile and the hydrodynamical obstruction.

\section{Dynamical scalar field simulations}
\label{sec:dyn}
We now turn to the dynamical simulation of the system. As opposed to the simulations studied in the previous section, where the wall thickness and velocity were imposed \emph{a priori} and ignored the plasma backreaction, we now implement a code which dynamically evolves both the fluid and the scalar field simultaneously.

\subsection{Analytical analysis of the early plasma evolution}

One of the main obstacles in solving Eq.~(\ref{eq:wallEOM1}) for a coupled field-plasma system is evaluating the time-dependent pressure $\mathcal{P}(t)$. As previously mentioned, due to the bubble propagation, the plasma develops sound waves that create a nontrivial temperature profile around the wall. We show here how to circumvent this problem by expanding the pressure around $t=0$. For the purpose of determining if hydrodynamic obstruction occurs, it will be sufficient to consider a planar wall geometry.

We start from Eqs.~\eqref{eq:EMTConservation}, setting $d=1$. Since the plasma is initially unperturbed, the temperature and velocity fields satisfy the initial conditions $T(0,z)=T_n$ and $v(0,z)=0$, with $z$ the coordinate perpendicular to the wall. We now expand the net pressure $\mathcal P(t)$ defined in Eq.~(\ref{eq:pressure_LTE}) in a Taylor series around $t=0$. We emphasize that, per se, this expansion is not justified up to the Jouguet velocity. However, we will observe that the results obtained here analytically are in close agreement (in some regions of the parameter space to be discussed) with the numerical results in Section \ref{sec:results}.

The first two time derivatives evaluated at $t=0$ are given by
\begin{subequations}
\begin{align}
    \left.\partial_t\mathcal P\right|_{t=0} 
    &= \left.\int\! dz\, \partial_z\phi\, \partial_t T\frac{\partial^2 V_{\rm eff}}{\partial\phi\partial T}\right|_{t=0} + \left.\int\! dz\, \partial_z\phi\, \partial_t \phi\frac{\partial^2 V_{\rm eff}}{\partial\phi\partial \phi}\right|_{t=0} = 0,\\
    \left.\partial_t^2\mathcal P\right|_{t=0} 
    &= \left.\int\! dz \,\partial_z\phi\,\partial_t^2 T\frac{\partial^2 V_{\rm eff}}{\partial\phi\partial T}\right|_{t=0} \, ,
\end{align}
\end{subequations} 
where we have used $\dot R(0)=0$, $\dot \phi(0) = 0$ and $\dot T(0)=0$. The latter follows directly from Eq.~(\ref{eq:EMTConservation}) evaluated at $t=0$.
The quantity $\ddot T(0)$ can be obtained by taking a time derivative of Eq.~(\ref{eq:energyConservation}) and evaluating the result at $t=0$. This yields
\begin{align}
    \ddot T(0) = \ddot R(0)\,\partial_z\phi\frac{\partial^2V_{\rm eff}}{\partial\phi\partial T}
    \left.\left(\frac{\partial^2V_{\rm eff}}{\partial T^2}\right)^{-1}\right|_{T=T_n}.
\end{align}
We used the fact that $\gamma =1$ and $v =0$ at $t=0$. 
Substituting this expression into the second derivative of the pressure and using the high-$T$ parametrization of the potential Eq.~(\ref{eq:bagPotential}) for concreteness, we obtain
\begin{subequations}
\begin{align}
\left.\partial_t^2\mathcal{P}\right|_{t=0} 
    &= \ddot R(0)\int_0^{\phi_0}\! d\phi\frac{2\phi(\phi_0-\phi)}{\phi_0 L}
    \left(\frac{\partial^2V_{\rm eff}}{\partial\phi\partial T}\right)^2
    \left.\left(\frac{\partial^2V_{\rm eff}}{\partial T^2}\right)^{-1}\right|_{T=T_n},\\
     &= -\frac{8(1-\Psi)^2 w_n}{3\phi_0^3L}\ddot R(0)\int_0^v\! d\phi\frac{\phi^3(\phi_0-\phi)}{\phi_0^2-(1-\Psi)\phi^2/3},
    \\
    &= -\ddot R(0)\left(\frac{2(1-\Psi)^2w_n}{15L}
    +\mathcal O((1-\Psi)^3)\right),
\end{align}
\end{subequations}
where in the first line we used the identity $dz\,\partial_z\phi=d\phi$ (which is valid at $t=0$), and in the second line we substituted Eqs.~(\ref{eq:bagPotential}) and~(\ref{eq:cCoeff}), using the following results 
\begin{equation}
    \frac{\partial^2V_{\rm eff}}{\partial T^2} = \frac{3w_n}{T_n^2}\bigg((1-\Psi)\frac{\phi^2}{3\phi_0^2} - \frac{T^2}{T_n^2}\bigg), \qquad \frac{\partial^2V_{\rm eff}}{\partial\phi\partial T} = \frac{2 (1-\Psi) w_n T\phi}{T_n^2 \phi_0^2}
\end{equation}
We have neglected the $c$ term in $\partial^2V_{\rm eff}/\partial T^2$, which is justified in the regime where $1-\Psi$ is small.

Finally, the Taylor expansion of the net pressure $\mathcal P(t)$ up to second order in time reads\footnote{It can be shown that the cubic term vanishes and that the quartic term gives a subleading contribution to the final result of this subsection for small $\alpha_p$ and $1-\Psi$. Therefore, this expansion is justified for weak PTs.}
\begin{align}\label{eq:expandedPressure}
    \mathcal{P}(t) = \frac{3\alpha_pw_n}{\nu}-\ddot R(0)\frac{(1-\Psi)^2w_n}{15L}t^2+\mathcal O(t^4).
\end{align}
From Eq.~(\ref{eq:wallEOM1}), one also obtains
\begin{align}
    \ddot R(0) &= \frac{\mathcal{P}(0)}{\sigma} = \frac{9L\alpha_p w_n}{\nu \phi_0^2}.
\end{align}
Neglecting the $1/R$ term and using the expanded pressure (\ref{eq:expandedPressure}), Eq.~(\ref{eq:wallEOM1}) has the solution
\begin{align}\label{eq:wallVelocity}
    \xi_w(t) = \partial_tR(t)=\left[1+\left(\frac{6\sigma}{6\mathcal{P}(0)t+\partial_t^2\mathcal{P}(0)t^3}\right)^2\right]^{-1/2} \, . 
\end{align}

Notice that the term $\partial_t^2\mathcal{P}(0)$ is negative and competes with the first positive term $\mathcal{P}(0)$. When they cancel each these two terms lead to a terminal velocity. Again, the expression Eq.~\eqref{eq:wallVelocity} is only expected to be reliable during the very early stages of bubble propagation. As the wall accelerates and approaches the speed of sound, nonlinear hydrodynamic effects become important and the expansion in Eq.~(\ref{eq:expandedPressure}) ceases to be valid. Nevertheless, this result provides valuable insight into the hierarchy of time scales governing the dynamics. In particular, it quantifies how rapidly the plasma responds to the presence of the wall and generates a backreaction force that acts to slow down its acceleration.

A useful diagnostic for assessing the onset of hydrodynamic obstruction is the maximum wall velocity predicted by Eq.~(\ref{eq:wallVelocity}), which is given by
\begin{align}\label{eq:vwMax}
    \xi_w^{\rm max} = \left(1-\frac{9\sigma^2\partial_t^2\mathcal{P}(0)}{8\mathcal{P}^{3}(0)}\right)^{-1/2}=\left(1+\frac{\nu^2\phi_0^2(1-\Psi)^2}{180L^2\alpha_p^2 w_n}\right)^{-1/2}
\end{align}
A large value of $\xi_w^{\rm max}$ indicates that the plasma responds too slowly to generate a significant friction force. In this regime, even if a SS deflagration or hybrid solution exists, it is likely to be dynamically bypassed, and the wall will instead evolve toward a detonation. Conversely, a small value of $\xi_w^{\rm max}$ implies that the plasma rapidly builds up sufficient friction to halt the wall acceleration, in which case the SS solution is expected to be dynamically realized.

To estimate the wall thickness $L$, we impose that a different moment of the equation of motion, corresponding to the pressure gradient across the wall, vanishes at $t=0$. This condition reads
\begin{align}
    0 &= \int\! dz\,\partial_z\phi\, \partial_z^2\phi
    \left.\left(\partial_t^2\phi-\partial_z^2\phi+\frac{\partial V_{\rm eff}}{\partial \phi}\right)\right|_{t=0}\nn
    &= \frac{8\phi_0^3}{105L^4}-\frac{12\phi_0\alpha_p w_n}{35\nu L^2(1-2b)}.
\end{align}
Solving this equation yields
\begin{align}\label{eq:width}
    L &= \frac{2\phi_0}{3}\sqrt{\frac{\nu(1-2b)}{2\alpha_p w_n}}\nn
    &\approx \left(\frac{\nu}{2\alpha_p w_n}\right)^{2/3}
    \left(\frac{2\phi_0}{3}\right)^{2}
    \left(\frac{\pi}{S_3}\right)^{1/3},
\end{align}
with $S_3$ the $\mathcal{O}(3)$-symmetric bounce action at the nucleation temperature. The second line follows from using the thin wall value of $b$ given in Eq.~(\ref{eq:twB}).

Finally, substituting Eq.~(\ref{eq:width}) into Eq.~(\ref{eq:vwMax}) and solving for $\alpha_p$ gives
\begin{subequations}
\begin{align}\label{eq:runawayCriterion}
    \alpha_{p,{\rm max}}^{\rm dyn} &=
    \frac{27\nu S_3}{640\pi}
    \sqrt{\frac{w_n}{5}}
    \left(\frac{\gamma_w^{\rm max}\xi_w^{\rm max}(1-\Psi)}{\phi_0}\right)^3\\
    &\approx60.64\left(\frac{1-\Psi}{\phi_0/T_n}\right)^3\left(\frac{\nu}{4}\right)\left(\frac{S_3/T_n}{140}\right)\left(\frac{g_+^*}{106.75}\right)^{1/2}\left(\frac{\gamma_w^{\rm max}\xi_w^{\rm max}}{1.381}\right)^3,
\end{align}
\end{subequations}
with $\gamma_w^{\rm max}=1/\sqrt{1-(\xi_w^{\rm max})^2}$. If $\xi_w^{\rm max}$ is interpreted as the maximal velocity allowed by the early time solution in Eq.~(\ref{eq:wallVelocity}) before the wall transitions to a detonation, then $\alpha_{p,{\rm max}}^{\rm dyn}$ represents the largest value of $\alpha_p$ compatible with the realization of a deflagration or hybrid solution. A priori, $\xi_w^{\rm max}$ could be a general function of the model parameters. We will see in the next subsection that it is very well approximated by a constant, with the optimal value given by $\xi_w^{\rm max}\approx0.81$.\footnote{Following this logic, $\xi_w^{\rm max}$ should then be interpreted as a numerical fitting parameter, and not as a physical velocity.} 

However, this bound is only valid for relatively weak phase transitions. For larger values of $1-\Psi$, the steady-state criterion Eq.~(\ref{LTE:criterion}) can become smaller than $\alpha_{p,{\rm max}}^{\rm dyn}$ and therefore gives a stronger bound on $\alpha_p$.\footnote{Note that the value of $\alpha_{p,{\rm max}}^{\rm SS}$ presented in Eq.~(\ref{LTE:criterion}) was computed with the template model with $c_{s,-}^2=\Psi/(2+\Psi)$ and $c_{s,+}^2=1/3$, which are the sound speeds of the potential (\ref{eq:bagPotential}) used in this paper. In general, the value of $\alpha_{p,{\rm max}}^{\rm SS}$ used in Eq.~(\ref{eq:finalCriterion}) should be computed with the EoS relevant to the potential used, and it may therefore differ from Eq.~(\ref{LTE:criterion}).}
In general, the maximal $\alpha_p$ is then given by
\begin{align}\label{eq:finalCriterion}
    \alpha_p^{\rm max} = \min\bigg[\alpha_{p,{\rm max}}^{\rm dyn},\alpha_{p,{\rm max}}^{\rm SS}\bigg].
\end{align}

Physically, this represents the fact that, for small plasma-wall coupling $1-\Psi$, the shock wave forms slowly and the dominant criterion $\alpha_{p,{\rm max}}^{\rm dyn}$ comes from comparing the bubble's acceleration to the shock formation time scale. For stronger plasma-wall coupling, the shock wave always has enough time to fully form before the wall reaches the shock front. Therefore, the relevant criterion is to determine if the shock wave can be large enough to stop the wall, which is quantified by $\alpha_{p,{\rm max}}^{\rm SS}$.

\subsection{Numerical results of the dynamical evolution of the fluid+scalar system}
\label{sec:results}

We finally come to the presentation of the results of our numerical analysis. To test the formula derived in the previous subsection, we performed a parameter scan using as a broad model the high-$T$ potential Eq.~(\ref{eq:bagPotential}). Using the procedure described in Appendix~\ref{sec:potential} and fixing $S_3/T_n=140$ and $g_+^*=106.75$ throughout, this potential depends on only three parameters, $\alpha_p$, $\Psi$, and $v/T_n$, which we sample randomly over the ranges
\begin{align}
    \alpha_p&\in[10^{-5}, 0.33],\nn
    1-\Psi&\in [0.01, 0.5],\nn
    \phi_0/T_n&\in [0.5,2].\nonumber
\end{align}
For each point sampled in this scan, we evolve the initial configuration by numerically solving the fluid equation~(\ref{eq:EMTConservation}) and the EoM~(\ref{eq:wallEOM}) using the algorithm described in Appendix~\ref{sec:numericalAlgorithm}. The initial condition is taken to be
\begin{subequations}\label{eq:initialCondition}
\begin{align}
    \phi(0,r)&= \phi_b(r)+\delta\phi(r),\\
    \partial_t\phi(0,r) &= \kappa\delta\phi(r),\\
    T(0,r) &= T_n,\\
    v(0,r)&=0,
\end{align}
\end{subequations}
where $\phi_b(r)$ is the bounce solution satisfying 
\begin{align}\label{eq:euclideanBounce}
    0=\partial_r^2\phi_b+\frac{2}{r}\partial_r\phi_b-V'(\phi_b,T_n),
\end{align}
and $\kappa$ and $\delta\phi(r)$ are the solutions to the eigenproblem
\begin{align}\label{eq:linearEOM}
    \kappa^2\delta\phi = \left[\partial_r^2+\frac{2}{r}\partial_r-V''(\phi_b,T_n)\right]\delta\phi,
\end{align}
with $\kappa>0$. The initial condition~(\ref{eq:initialCondition}) is obtained by linearizing the EoM around the bounce solution $\phi_b$ and integrating forward from $t=-\infty$ to $t=0$. Eq.~(\ref{eq:linearEOM}) follows from writing the solution of the linearized EoM as
\begin{align}
    \phi_{\rm lin}(t,r) = \phi_b(r)+\delta\phi(r)e^{\kappa t}.
\end{align}
Note that the overall magnitude of $\delta\phi$ is arbitrary; we choose it small enough to remain well within the linear regime. In practice, $\phi_b$ is computed using the package {\tt CosmoTransitions}~\cite{Wainwright:2011kj}, while $\kappa$ and $\delta\phi$ are computed by discretizing Eq.~\eqref{eq:linearEOM} with the pseudospectral method presented in Appendix \ref{sec:numericalAlgorithm} to transform it in a normal matrix eigenvalue problem.\footnote{Alternatively, $\delta\phi\approx \partial_r\phi$ in the thin-wall limit. Also, $\kappa$ can be computed with the package {\tt BubbleDet}~\cite{Ekstedt:2023sqc}.}

The subsequent evolution of the field profile according to Eqs.~(\ref{eq:EMTConservation}) and~(\ref{eq:wallEOM}) falls into two distinct classes of final configurations. If the initial acceleration is small and the wall-plasma coupling is large, the wall eventually reaches a subjouguet terminal velocity, and the 
field and plasma profiles converge to the expected self-similar solution. Otherwise, the wall accelerates beyond the self-similar prediction and runs away toward $\xi_w\to 1$. This occurs when the hydrodynamic obstruction does not develop fast enough before the wall overtakes the shock front.

\begin{figure}
    \centering
    $\alpha_p=0.002<\alpha_p^{\rm max}$\hspace{0.33\linewidth}$\alpha_p=0.008>\alpha_p^{\rm max}$\\
    \includegraphics[width=0.48\linewidth]{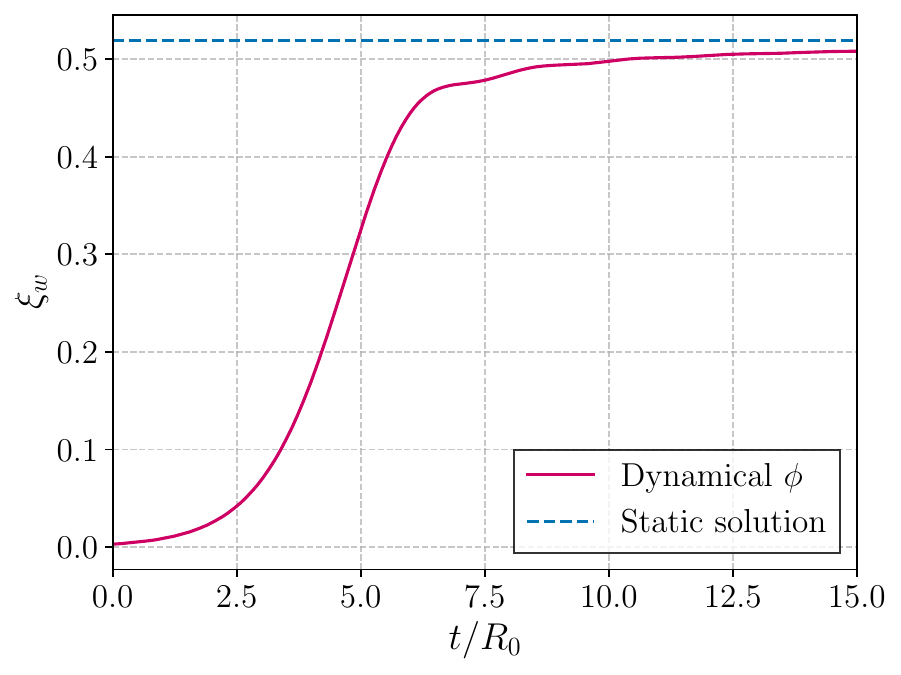} \hspace{0.03\linewidth}\includegraphics[width=0.48\linewidth]{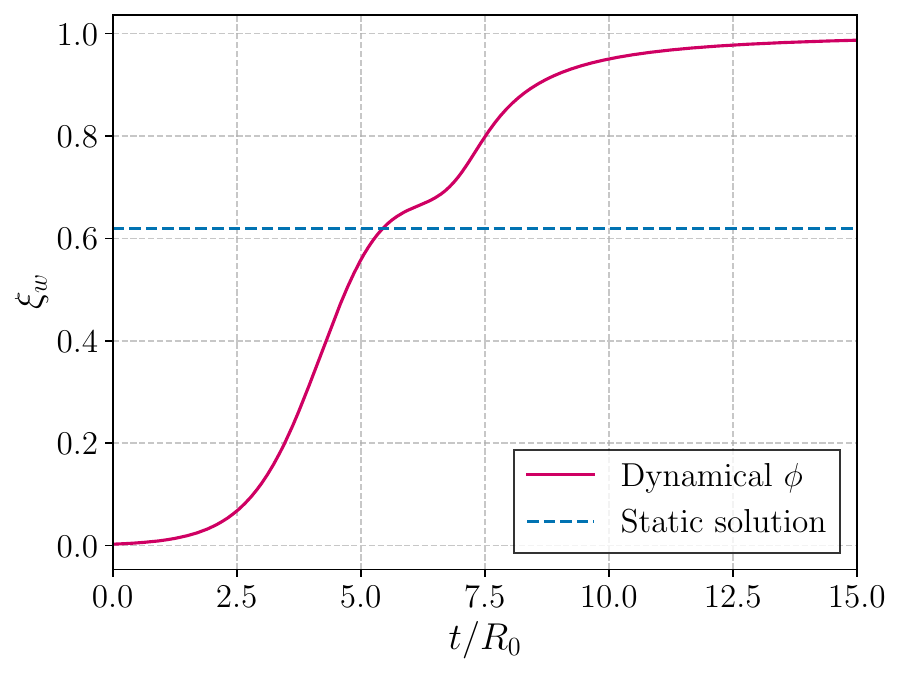}\\
    \includegraphics[width=0.48\linewidth]{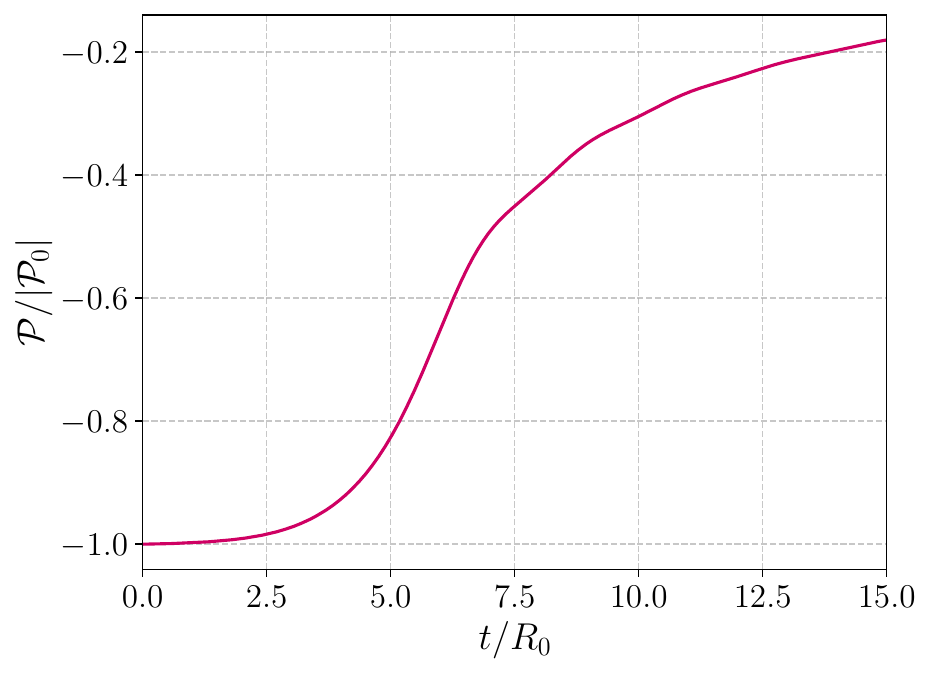} \hspace{0.03\linewidth}\includegraphics[width=0.48\linewidth]{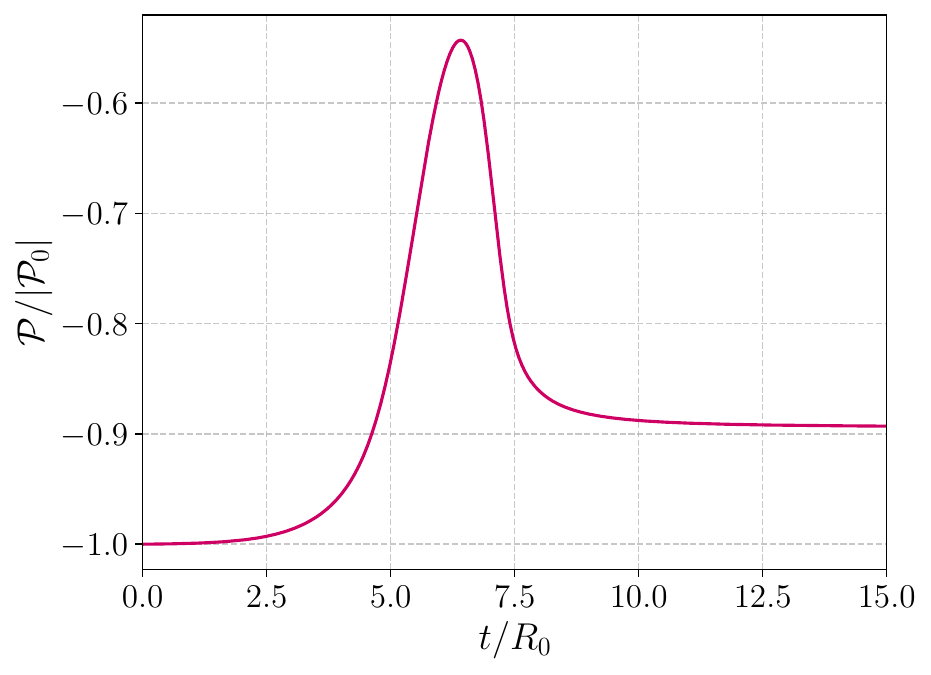}\\
    \includegraphics[width=0.48\linewidth]{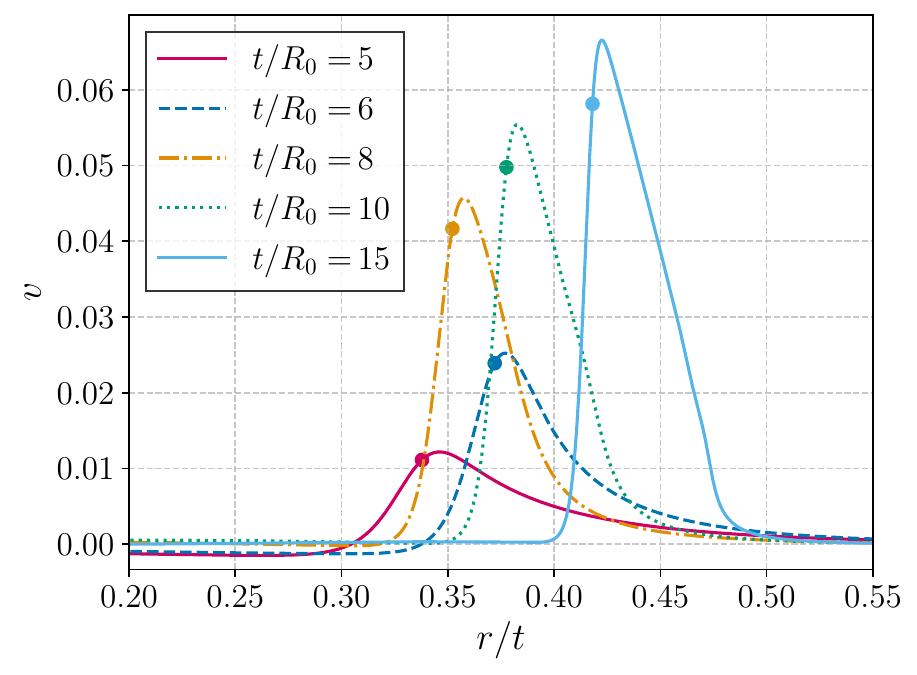} \hspace{0.03\linewidth}\includegraphics[width=0.48\linewidth]{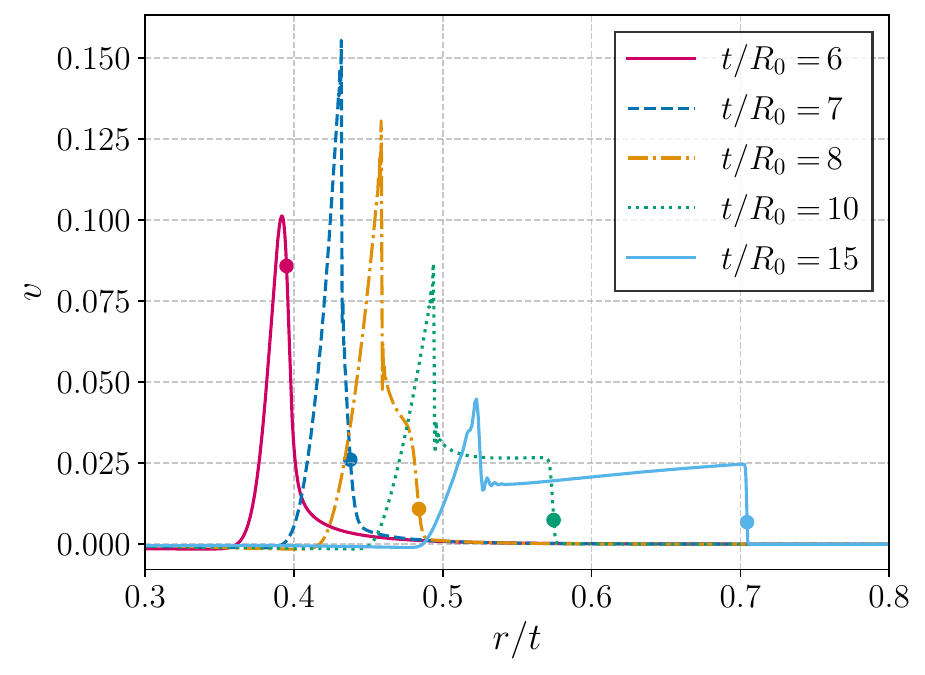}
    \caption{Examples of solutions with $\psi=0.95$ and $\langle\phi\rangle/T_n = 1$ 
($\alpha_p^{\rm max}\approx0.0071$). \textbf{Left panel}: Solution for 
$\alpha_p=0.002<\alpha_p^{\rm max}$, which experiences hydrodynamic 
obstruction and reaches a subjouguet terminal velocity. \textbf{Right panel}: Solution 
for $\alpha_p=0.008>\alpha_p^{\rm max}$, which accelerates beyond its 
expected self-similar solution and becomes a runaway. The top row shows 
the wall velocity as a function of time together with the expected 
terminal velocity. The middle row shows the pressure on the wall as a function of time. The bottom row shows the plasma velocity profile at 
several times during the bubble evolution, with the dots showing the position of the wall's center. $R_0$ is the initial bubble radius.}
    \label{fig:exampleSolutions}
\end{figure}

\begin{figure}
    \centering
    \includegraphics[width=0.5\linewidth]{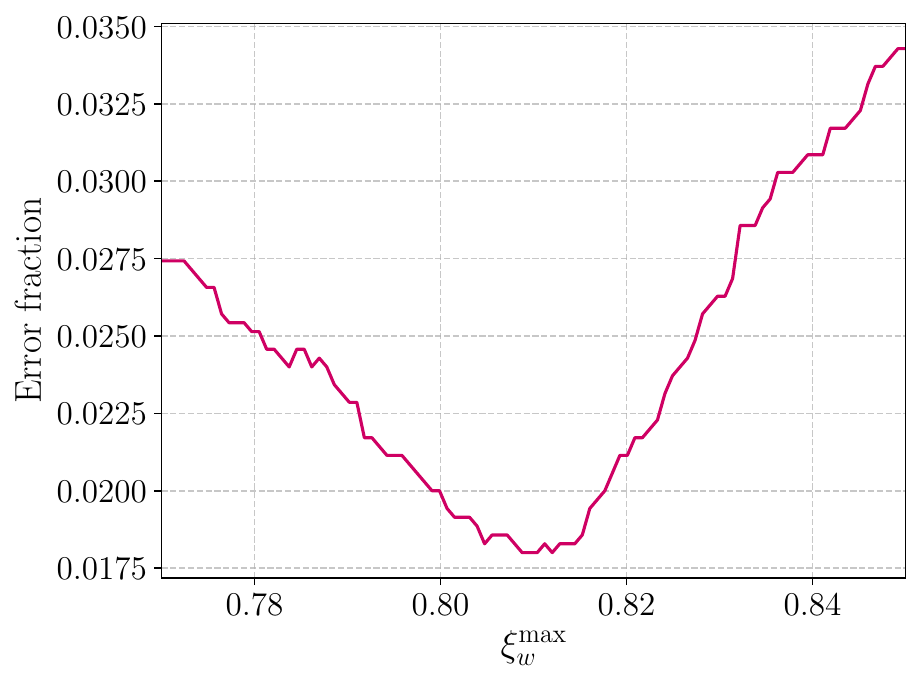}
    \caption{Fraction of points classified incorrectly as a function of the fitting parameter $\xi_w^{\rm max}$. The best fit is given by $\xi_w^{\rm max}\simeq 0.81$.}
    \label{fig:bestFit}
\end{figure}

Representative examples of the field profile and wall velocity for both classes of solutions are shown in Fig.~\ref{fig:exampleSolutions}. The left column shows a solution with $\alpha_p<\alpha_p^{\rm max}$ that 
experiences hydrodynamic obstruction. After a brief initial acceleration, 
the backreaction from the plasma grows until the pressure approaches zero and the wall velocity reaches the 
value predicted by the static analysis. In the bottom-left panel, one can 
observe that the shock wave develops rapidly, generating the pressure 
necessary to halt the wall. Importantly, the wall remains behind the shock 
front at all times.

The second class of solution ($\alpha_p>\alpha_p^{\rm max}$) is shown in 
the right column. Although the wall acceleration does decrease upon 
reaching the expected terminal velocity, it does not do so quickly enough 
to stop the wall completely, and the acceleration eventually begins 
increasing again when the pressure goes back down to a constant negative value. One can see in the bottom-right panel that this 
coincides approximately with the moment when the wall (whose position is shown by the dots) catches up with the 
shock front and overtakes it ($t/R_0\sim 6\ \mathrm{to}\ 8$).  This 
reduces the backreaction pressure on the wall, which is then free to 
accelerate beyond the Jouguet velocity toward the speed of light. We observe that even after the wall crossed the Jouguet velocity, a remnant of the shock remains behind the wall, but in the absence of source, it quickly decays. 

In Eq.\eqref{eq:runawayCriterion}, there is still one remaining parameter $\xi_w^{\rm max}$ that needs to be fitted against this scan in order to have a full determination of $\alpha_p^{\rm max}$. Figure \ref{fig:bestFit} shows the fraction of points classified incorrectly as a function of this parameter. The optimal value of $\xi_w^{\rm max}$ which minimizes the number of errors is $\xi_w^{\rm max} \simeq 0.81$. This may seem quite fast, as the shock front travels at the speed of sound for weak transitions, so the wall can seemingly travel much faster than the shock front before becoming a detonation. The main reason for this high value is that what actually matters is not the relative speed between the wall and the shock front, but rather their relative position. Since the wall requires some time to accelerate, while the shock front always travels at the speed of sound, it takes some time for the wall to catch up with the shock front even after it becomes supersonic.

The main result of the scan is shown in Fig.~\ref{fig:scatter_plot}, which shows for each point sampled the value of the runaway criterion (\ref{eq:finalCriterion}) as a function of $\alpha_p$. Red points are runaway solutions, while the blue points are solutions that reach a subjouguet terminal velocity. Furthermore, the dots represent weak phase transitions where $\alpha_p^{\rm max}$ is determined by $\alpha_{p,{\rm max}}^{\rm dyn}$, while crosses are determined by $\alpha_{p,{\rm max}}^{\rm SS}$. The simple criterion $\alpha_p^{\rm max}$ works surprisingly well for a large range of parameters, leading to an error rate of only 1.8\% (see the best fit point of Fig.~\ref{fig:bestFit}). This confirms that, although the derivation of $\alpha_p^{\rm max}$ 
presented in the previous subsection relies on rough approximations and 
cannot quantitatively predict the wall evolution, it does accurately 
capture how the timescales for wall acceleration and for building up the 
backreaction pressure vary with the model parameters. It is therefore a 
valuable tool for analyzing the early evolution of the bubble.

Furthermore, we verified that the static analysis and the dynamical simulations agree well on the terminal velocity $\xi_w$ whenever both methods predict a subjouguet velocity (the blue points in Fig.~\ref{fig:scatter_plot}). This is illustrated, for example, in the upper-left corner of Fig.~\ref{fig:exampleSolutions}, where the dynamical velocity converges to the static solution. Across the entire scan, the average discrepancy between the static and dynamical predictions was always within our numerical tolerance.

\section{Conclusion}
\label{sec:conclusion}
Our work challenges the conventional assumption that bubble wall dynamics during first-order phase transitions can be accurately and completely described by steady-state hydrodynamics. By analyzing the transient evolution of the pressure on the bubble wall, we demonstrate that the formation of the heating shock wave, which is a critical component of hydrodynamical obstruction, often occurs on timescales comparable to or longer than the wall’s acceleration phase. This delay undermines the efficacy of steady-state predictions, particularly near the Jouguet velocity, where obstruction is traditionally assumed to be maximal.

Our key result, Eq.~\eqref{eq:main_result}, provides a revised criterion for the maximal driving pressure $\alpha_{p}^{\rm max}$ that separates deflagration/hybrid regimes from detonations/runaway walls. Unlike the steady-state LTE prediction (\ref{LTE:criterion}), this criterion accounts for the finite timescale of plasma response, revealing that hydrodynamic obstruction is less restrictive than previously thought. Numerical simulations confirm that the pressure on the wall saturates rapidly for supersonic velocities but may take significantly longer for subsonic or near-Jouguet speeds, aligning with the observed separation of regimes in Fig.~\ref{fig:scatter_plot}.

The dynamical scalar field simulations further validate our analytical framework, showing that early-time pressure evolution can be approximated via Taylor expansion, while late-time behavior requires full hydrodynamic treatment. These findings have critical implications for gravitational wave spectra and baryogenesis, as they relax constraints on phase transition parameters (e.g., $\Psi < 0.77$) previously deemed incompatible with runaway walls from a pure steady state analysis\cite{Ai:2024shx} (at least for $c_{s,+}^2=c_{s,-}^2= 1/3$). We have also verified that, in the specific case of a terminal wall velocity, the velocity computed from the dynamical simulations agrees very well with the pure ``steady state'' velocity, with a discrepancy always lying within the imposed numerical tolerance. 

This last observation confirms that static analysis remains a valuable tool for determining the terminal wall velocity, although it must be supplemented by a dynamical analysis (or, more prosaically, by the criterion in Eq.~\eqref{eq:main_result}) to establish whether hydrodynamic obstruction occurs. We therefore suggest computing $\xi_w$ in two steps: first, determine which type of solution is physically relevant using Eq.~\eqref{eq:main_result}; then, if hydrodynamic obstruction occurs, compute $\xi_w$ with standard SS methods.

\paragraph{Limitations of the current analysis}

Finally, our analysis neglected a few possibly relevant effects:
\begin{itemize}
    \item \textbf{Friction from the entropy production}: Friction from the dissipation of entropy is unavoidable also during the early stage of the bubble expansion. Contrary to the LTE backreaction, this dissipative friction does not require the formation of a shock wave to be effective. We therefore expect these effects to significantly increase $\alpha_p^{\rm max}$ as the additional pressure on the wall will slow down its acceleration. We leave the inclusion of entropy production to following studies (Notice that this has been studied in \cite{Krajewski:2024zxg}).
    \item \textbf{Effects of non-spherical effects in nucleation}: The early shape, size and acceleration of the bubble are not properly described by the classical equations of motion which we have solved in this paper, they are actually better described by a Langevin-like equation. In this context, it was found that the very early acceleration of the bubble could be slower than the one obtained from classical numerical solutions\cite{Gould:2024chm}. 
    \item \textbf{Timescale of thermalisation}: Our approach assumed instantaneous thermalisation ahead of the bubble front, particularly in the compression wave. In the context of the EWPT, this assumption is likely justified for colored particles, as the quark-gluon plasma thermalises over very short time scales of order $\sim 1/T$. However, the thermalisation of weakly interacting particles is expected to take longer and would require a more detailed analysis. For BSM or dark phase transitions, our analysis should remain valid for strongly coupled systems. A more complete treatment would entail solving the full set of Boltzmann equations also ahead of the bubble front.
\end{itemize}

We hope to be able to explore these effects in future studies.

\section*{Acknowledgments}

We are pleased to thank Xander Nagels and Simone Blasi for collaboration and discussions in the early stages of this project. We also thank Xander Nagels, Ignacy Nałęcz, Wenyuan Ai, Alex Azatov for comments on the manuscript. BL would also like to thank the University of Barcelona for their hospitality.

Research at Perimeter Institute is supported in part by the Government of Canada through the Department of Innovation, Science and Economic Development Canada and by the Province of Ontario through the Ministry of Colleges and Universities.

 M.V. is funded by the European Union (ERC, HoloGW, Grant Agreement No. 101141909). M.V. also acknowledges financial support from Grant CEX2024-001451-M funded by 
 \\
 MICIU/AEI/10.13039/501100011033, from Grant No. PID2022-136224NB-C22 from the Spanish Ministry of Science, Innovation and Universities, and from Grant No. 2021-SGR-872 funded by the Catalan Government.
\appendix

\section{Numerical algorithm}
\label{sec:numericalAlgorithm}

We describe here the numerical algorithm used to solve the conservation equations~(\ref{eq:EMTConservation}) and the EoM~(\ref{eq:wallEOM}). We use a pseudospectral method similar to those of Refs.~\cite{Laurent:2022jrs} and \cite{Ekstedt:2024fyq}, which computed the wall velocity in a static calculation. A more detailed account of pseudospectral methods can be found in Ref.~\cite{boyd01:CFS}.

The method discretizes the spatial coordinate by expanding the unknown functions $\phi(t,r)$, $T(t,r)$, and $v(t,r)$ in a series of Chebyshev polynomials and requiring the equations to be satisfied exactly on a prescribed grid. For the scalar field profile, we write
\begin{align}\label{eq:chebyshevSeries}
    \phi(t,r)\approx\phi_N(t,r)=\sum_{n=0}^{N-1} a_n(t)T_n(\chi(r)),
\end{align}
where $N$ is the number of grid points and $\chi$ is a new coordinate mapping the radial coordinate onto the interval $\chi\in[-1,1]$ on which the Chebyshev polynomials $T_n$ are defined.

Numerically, it is advantageous to choose $\chi(r)$ so that $\phi(t,r(\chi))$ is as smooth as possible when expressed in terms of $\chi$. We adopt the adaptive mapping
\begin{align}
    \chi(r) = \frac{r^2 - R^2(t)}{\sqrt{(r^2 - R^2(t))^2 + L_\chi^2(r - r_{\rm min})^2}},
\end{align}
where $R(t)$ tracks the position of the wall, $L_\chi$ is the length scale over which the solution is expected to be nontrivial, and $r_{\rm min}$ is the smallest value of $r$ considered. As $r$ varies from $r_{\rm min}$ to $\infty$, $\chi$ increases monotonically from $-1$ to $1$. In principle, $r_{\rm min}$ should be set to $0$ to cover the entire volume; here we instead assign it a small positive value to avoid the coordinate singularity at the origin. We have verified that our results are insensitive to $L_\chi$ and $r_{\rm min}$, which we fix to $L_\chi=R_0$ and $r_{\rm min}=R_0/100$.

Expanding in a basis of Chebyshev polynomials offers two advantages. First, Eq.~(\ref{eq:chebyshevSeries}) can be recast as a cosine series through the relation $T_n(\cos{\theta})=\cos{n\theta}$, allowing us to use the fast cosine transform for operations on $\phi$ such as differentiation. Second, if all derivatives of $\phi$ are continuous over the whole domain, the coefficients $a_n$ decay as $e^{-cn}$ for sufficiently large $n$, with $c$ a positive constant. The truncation error is therefore expected to be exponentially small,
\begin{align}
    |\phi-\phi_N|\sim e^{-cN},\qquad {\rm large\ }N,
\end{align}
which is far superior to finite difference schemes, whose error scales only as $1/N^{c}$.

In practice, $N$ conditions are required to fix the $N$ coefficients $a_n$. A simple prescription is to demand that the residual $R[\phi]$ of the equation to be solved be orthogonal to every basis element $T_n$:
\begin{align}\label{eq:galerkin}
    0=\int_{-1}^1 d\chi\, w(\chi)T_n(\chi)R[\phi(\chi)],\qquad n=0,\cdots,N-1,
\end{align}
with $w(\chi)=1/\sqrt{1-\chi^2}$ the weight associated with the Chebyshev polynomials. This is the \emph{Galerkin} method. While conceptually clear, it becomes tedious to implement when $R[\phi]$ has nonconstant coefficients or is nonlinear in $\phi$, since the integrals in Eq.~(\ref{eq:galerkin}) may then lack a closed form. This is the case, for instance, for the scalar EoM, which contains a nonlinear potential, $R[\phi]\supset V_{\rm eff}(\phi)$.

A simpler alternative is the \emph{pseudospectral} method, in which the integral in Eq.~(\ref{eq:galerkin}) is evaluated by Gaussian quadrature. One can then show that this is equivalent to requiring the equation to be satisfied exactly on the grid $\chi_i$ formed by the quadrature nodes,
\begin{align}
    R[\phi(\chi_i)]=0,\qquad i=0,\cdots,N-1.
\end{align}
We use the Gauss-Lobatto grid, which includes the endpoints and the extrema of $T_{N-1}$,
\begin{align}
    \chi_i = \cos{\bigg(\frac{\pi (N-i-1)}{N-1}\bigg)}.
\end{align}
With this discretization, the scalar EoM can be written in the field basis $\phi_i(t)\equiv\phi(t,\chi_i)$ as
\begin{align}
    \partial_t^2\phi_i=\partial_r^2\phi_i+\frac{2}{r}\partial_r\phi_i-\frac{\partial V_{\rm eff}}{\partial\phi}(\phi_i,T_i),\qquad i=0,\cdots, N-1,
\end{align}
and each component $\phi_i(t)$ can be integrated forward in time with standard methods such as Runge-Kutta integration.

In principle, the derivative $\partial_r\phi_i$ can be expressed in matrix form as $\partial_r\phi_i=\sum_jD_{ij}\phi_j$ (and similarly for $\partial_r^2\phi_i$), a slow $\mathcal{O}(N^2)$ operation because $D$ is dense. It is more efficient, however, to differentiate in the spectral basis $a_n$, where recursion relations yield the derivative in only $\mathcal{O}(N)$ operations. As noted above, the properties of the Chebyshev polynomials allow one to pass from the field basis $\phi_i$ to the spectral basis $a_n$ in $\mathcal{O}(N\log_2 N)$ operations using the fast cosine transform. Both $\partial_r\phi_i$ and $\partial_r^2\phi_i$ can therefore be computed in $\mathcal{O}(N\log_2 N)$ operations by first going to the spectral basis, taking the derivative, and finally going back to the field basis. Note that the same method can be applied to the conservation equations (\ref{eq:EMTConservation}) and the functions $T(t,r)$ and $v(t,r)$ in a completely analogous way.

\begin{figure}
    \centering
    \includegraphics[width=0.5\linewidth]{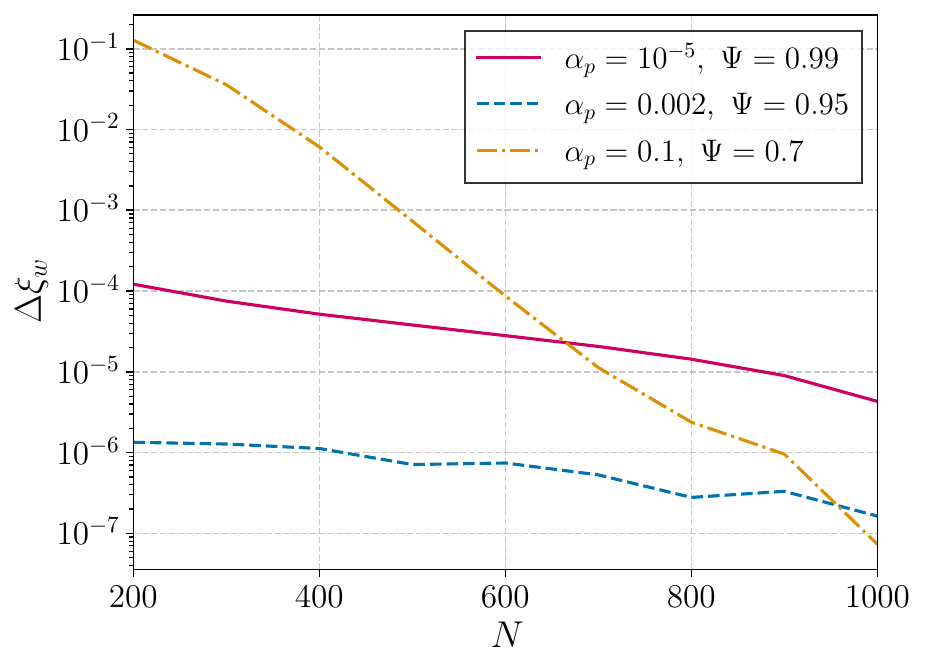}
    \caption{Error in the terminal wall velocity as a function of the number of grid points $N$.}
    \label{fig:convergence}
\end{figure}

To confirm that the algorithm converges as expected, Fig.~\ref{fig:convergence} shows the error in the terminal velocity $\Delta\xi_w\equiv|\xi_w(N)-\xi_w(N\to\infty)|$ as a function of the number of grid points $N$ for three choices of parameters. The error is consistent with exponential convergence, $\Delta\xi_w\sim e^{-cN}$, although the rate $c$ is quite model-dependent. Notably, the strongest PT has the largest error at small $N$, but also the largest rate $c$, so that it eventually becomes the most precise. To keep all our calculations accurate, we set $N=1000$ throughout the remainder of this paper, which should ensure that our solutions have converged.

\section{Higgs potential and nucleation}
\label{sec:potential}

Although fixed-scalar simulations can be performed without specifying a complete finite-temperature potential $V_{\rm eff}(\phi,T)$, several quantities of physical interest, notably the frictional pressure exerted on the bubble wall, depend nontrivially on the detailed form of $V_{\rm eff}(\phi,T)$. Moreover, dynamical simulations in which the scalar field profile $\phi(t, r)$ is obtained by explicitly solving its equation of motion necessarily require a full specification of the potential. As a consequence, unlike static determinations of the wall velocity $\xi_w$, which depend only on the parameters $\alpha_{p}$ and $\Psi$ when the bag EoS is assumed, time-dependent simulations of $\xi_w$ are intrinsically more model-dependent, as they also probe the functional form of $V_{\rm eff}(\phi,T)$.

In this section, we introduce a simplified potential $V_{\rm HT}(\phi,T)$ whose purpose is to provide a model-independent parametrization that captures the physical features of a general potential $V_{\rm eff}(\phi,T)$ that are relevant for the dynamics of the bubble wall. In particular, we construct $V_{\rm HT}(\phi,T)$ to correspond to the simplest high-$T$ expanded potential that is phenomenologically viable while reproducing the correct values of $\alpha_{p}$ and $\Psi$ as those obtained from the underlying potential $V_{\rm eff}(\phi,T)$. By construction, $V_{\rm HT}(\phi,T)$ should therefore be understood as an effective description that encodes these macroscopic thermodynamic properties, rather than as a controlled approximation to the true scalar potential.

To this end, we choose $V_{\rm HT}(\phi,T)$ to be a quartic polynomial in $\phi$ with thermal corrections proportional to $T^4$ and $\phi^2 T^2$. Physically, these correspond to the contributions from the relativistic thermal pressure and thermal masses, respectively. We further require the potential to possess two local minima located at $\langle\phi\rangle_{\rm false}=0$ and $\langle\phi\rangle_{\rm true}=\phi_0$ when $T=T_n$. A simple potential satisfying these requirements is given by
\bea\label{eq:bagPotential}
V_{\rm HT}(\phi,T)
=
\frac{a}{12}\phi^2
\left[6\phi_0^2b-4\phi_0(1+b)\phi+3\phi^2\right]
-\frac{w_n}{4}\frac{T^4}{T_n^4}
+\frac{c}{2}(T^2-T_n^2)\phi^2,
\eea
where
\bea
w_n=\frac{2\pi^2}{45}g_+^\star T_n^4
\eea
denotes the enthalpy density in front of the bubble, and $g_+^\star$ is the corresponding effective number of relativistic degrees of freedom.
The dimensionless parameters $a$, $b$ and $c$ are fixed by matching the macroscopic phase transition parameters $\alpha_{p}$ and $\Psi$ according to
\begin{subequations}\label{eq:bagParameters}
\begin{align}
a &= \frac{36\alpha_p(T_n)w_n}{\nu(1-2b)\phi_0^4},\\
\label{eq:cCoeff}
c &= \frac{(1-\Psi) w_n}{T_n^2\phi_0^2} \, ,
\end{align}
\end{subequations}
which we obtained by inverting the definitions
\begin{equation}
   \alpha_p \equiv \frac{\nu\Delta V}{3w_+}, \qquad \qquad  \Psi \equiv \frac{w_-}{w_+} \,  ,
\end{equation}
with $\nu=1+1/c_{s,-}^2$.\footnote{Note that $c_{s,-}^2=\frac{\Psi}{2+\Psi}$ for $V_{\rm HT}$, but we will keep the dependence on $c_{s,-}^2$ explicit to stay model-independent.}
Substituting the parameters in Eq.~(\ref{eq:bagParameters}) into Eq.~(\ref{eq:bagPotential}) leaves only the model parameters $b$ to be specified. It controls both the position and the height of the potential barrier, interpolating between a vanishing barrier for $b=0$ and an infinitely high barrier in the limit $b\to 1/2$. It therefore has a direct connection to the surface tension of the bubble wall, which can play a determining role in the early stages of bubble growth. We fix $b$ by requiring the tunneling action $S_3$ to take a prescribed value. Throughout this work, we adopt
\begin{align}
    \frac{S_3}{T_n}=140,
\end{align}
which is appropriate for a phase transition occurring at the electroweak scale.

For completeness, $S_3$ denotes the $O(3)$-symmetric Euclidean action,
\begin{align}
    S_3
    =
    4\pi\int dr\, r^2
    \left[
        \frac{1}{2}\big(\partial_r\phi_b\big)^2
        +V_{\rm HT}(\phi_b,T_n)
    \right],
\end{align}
where $\phi_b(r)$ is the bounce solution satisfying Eq.~\eqref{eq:euclideanBounce}.
In our numerical implementation, $S_3$ is evaluated numerically using the package {\tt CosmoTransitions} \cite{Wainwright:2011kj}. In some cases, it can be useful to have an analytic expression for $b$. This can be obtained with the thin-wall approximation, which corresponds here to the limit $b\to 1/2$. Following the method of Ref.~\cite{Matteini:2024xvg}, one quickly obtains,
\begin{align}
    S_3^{\rm tw}(T_n) &= \frac{32\pi \phi_0}{9\sqrt{a}}\frac{b^{3/2}(1+b)^2}{(1-2b)^2(2-b)^2} \nn
    &\approx \frac{8\pi \phi_0^3\sqrt{\nu}}{27\sqrt{2\alpha_p w_n}(1-2b)^{3/2}}, 
\end{align}
where we have expanded around $b=1/2$ to obtain the second line. This can finally be solved for $b$ as 
\begin{align}\label{eq:twB}
    b_{\rm tw}\approx\frac{1}{2}\left[1-\frac{1}{2}\left(\frac{\nu}{\alpha_p w_n}\right)^{1/3}\left(\frac{\phi_0}{3}\right)^2\left(\frac{16\pi}{S_3^{\rm tw}}\right)^{2/3}\right].
\end{align}

In summary, the potential $V_{\rm HT}(\phi,T)$ is fully specified by a small set of parameters with clear physical interpretations, namely the scalar vacuum expectation value $v$, the nucleation temperature $T_n$, the effective number of relativistic degrees of freedom in front of the wall $g_+^\star$, the phase transition strength $\alpha_p$, and the enthalpy ratio $\Psi$.

\section{Maximal $\alpha_{p}$ in static calculations}
\label{app:alphaMax}

We derive here the maximal value of $\alpha_p$ that allows for a hybrid solution to exist in a static calculation. In other words, we wish to find a function $\alpha_{p,{\rm max}}^{\rm SS}(\Psi)$ such that $\alpha_p<\alpha_{p,{\rm max}}^{\rm SS}(\Psi)$ guaranties the existence of a static deflagration or hybrid solution. For simplicity, we will assume that the plasma can be described accurately by the bag EoS, which allows us to set the speed of sound to $c_s=1/\sqrt{3}.$

As shown in Ref.~\cite{Ai:2023see}, a LTE static solution can be found by solving the matching equations
\begin{subequations}\label{eq:standardMatching}
\begin{align}
    \frac{v_+}{v_-} &= \frac{v_+v_--1/3+\alpha_{+}}{v_+v_--1/3+v_+v_-\alpha_{+}},\\
    12v_+v_-\alpha_{+} &= (1-3v_+v_-)\left(1-3\alpha_{+}-\left(\frac{\gamma_+}{\gamma_-}\right)^4\Psi\right).
\end{align}
\end{subequations}

We are looking for the strongest hybrid solution possible for a given value of $\Psi$. This coincides with the fastest hybrid wall, which implies $\xi_w=v_J$, with $v_J$ the Jouguet velocity
\begin{align}\label{eq:jouguet}
    v_J=\frac{1}{\sqrt{3}}\left(\frac{1+\sqrt{\alpha_{n}(2+3\alpha_{n})}}{1+\alpha_{n}}\right).
\end{align}
Furthermore, in this limiting case, the shock wave becomes infinitesimally thin which sets the additional constraint $\xi_w=\xi_{\rm sf}$, with $\xi_{\rm sf}$ the shock front velocity. The plasma velocity must also satisfy the following matching condition around the shock front
\begin{align}\label{eq:sfMatching}
    v_{\rm sf,+}v_{\rm sf,-} = 1/3.
\end{align}

Finally, hybrid solutions satisfy the boundary conditions $v_-=c_s=1/\sqrt{3}$, $v_{\rm sf,+}=\xi_{\rm sf}=\xi_w$ and $T_{\rm sf,+}=T_n$. Additional, the plasma in front can be related to the plasma behind the shock front by integrating the fluid equations across the shock wave. But since the shock wave has a vanishing thickness in this limit, the integration becomes trivial and one obtains $v_+=v_{\rm sf,-}$ and $T_+=T_{\rm sf,-}$.

We now have all the ingredients needed to determine what value of $\alpha_{n}$ corresponds to this strongest-hybrid limit. First, one can eliminate $\alpha_{+}$ from Eqs.~(\ref{eq:standardMatching}), which leads to a quartic equation in $v_+$ with the solution
\begin{align}\label{eq:vplus}
    v_+ = \frac{2}{\sqrt{3}}\sin\left(\frac{1}{3}\arcsin\Psi\right).
\end{align}
Using $v_{\rm sf,-}=v_+$ and $v_{\rm sf,+}=v_J$, one can also solve Eqs.~(\ref{eq:jouguet}) and (\ref{eq:sfMatching}) for $\alpha_{n}$ as
\begin{align}
    \alpha_{n}=\frac{1-2\sqrt{3}v_+ +3v_+}{9v_+ -1}.
\end{align}
Substituting Eq.~(\ref{eq:vplus}) in the last equation, we finally obtain
\begin{align}
    \alpha_{n} = \frac{\left(1-2\sin\left(\frac{1}{3}\arcsin\Psi\right)\right)^2}{5-6\cos\left(\frac{2}{3}\arcsin\Psi\right)}.
\end{align}

Using the relation $\alpha_{n}=\alpha_p+(1-\Psi)/3$, the maximal value of $\alpha_p$ that allows for a static hybrid solution is then
\begin{subequations}
\begin{align}
\label{eq:SS_alpha_max}
    \alpha_{p,{\rm max}}^{\rm SS, bag} &= \frac{\left(1-2\sin\left(\frac{1}{3}\arcsin\Psi\right)\right)^2}{5-6\cos\left(\frac{2}{3}\arcsin\Psi\right)} - \frac{1-\Psi}{3} \\
    \label{eq:alphaMaxExpandedApp}
    &= \frac{10}{9}\sqrt{\frac{2}{3}}(1-\Psi)^{3/2}+\mathcal O((1-\Psi)^2).
\end{align}
\end{subequations}
It is interesting to note that $\alpha_{p,{\rm max}}^{\rm SS,bag}$ diverges when $\Psi=\frac{4}{3\sqrt{3}}\approx 0.77$. Below this value of $\Psi$, it is always possible to find a static deflagration or hybrid solution no matter how strong the phase transition is.

\begin{figure}
    \centering
    \includegraphics[width=0.5\linewidth]{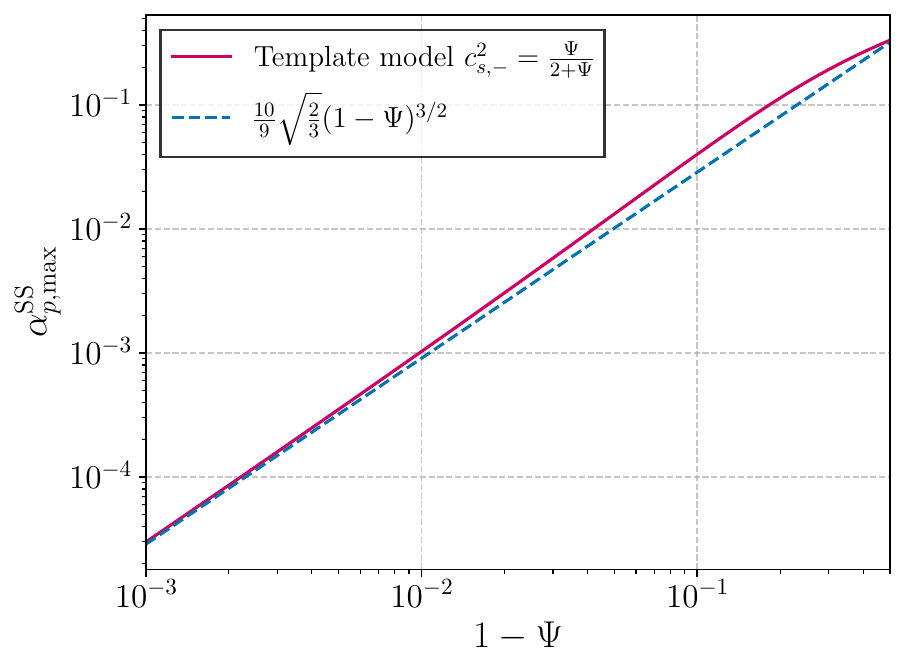}
    \caption{$\alpha_{p,{\rm max}}^{\rm SS}$ computed within the template model with $c_{s,+}^2=1/3$ and $c_{s,-}^2=\Psi/(2+\Psi)$, together with the small $1-\Psi$ expansion.}
    \label{fig:alphaMax}
\end{figure}

We emphasize that this result is valid within the bag EoS, where it is assumed that both the speeds of sound in front of and behind the wall are $c_{s,\pm}^2=1/3$. The high-$T$ potential (\ref{eq:bagPotential}), however, has $c_{s,+}^2=1/3$ and $c_{s,-}^2(T_n)=\Psi/(2+\Psi)$, with the latter also depending on the temperature in a nontrivial way. Neglecting this temperature dependence, this EoS can be represented by the template model \cite{Leitao:2014pda}. Within this EoS, the calculation of $\alpha_{p,{\rm max}}^{\rm SS}$ can be made similarly to what was presented here, although the equations now depend on $c_{s,-}^2(T_n)$ and must be solved numerically \cite{Ai:2023see}. We show in Fig.~\ref{fig:alphaMax} the result of this calculation with its small $1-\Psi$ expansion, which is the same as the bag EoS case (\ref{eq:alphaMaxExpandedApp}). We see that the expansion is numerically quite accurate for a large range of $\Psi$. Therefore, this is the value we adopt in this paper:
\begin{align}
    \alpha_{p,{\rm max}}^{\rm SS}\approx \frac{10}{9}\sqrt{\frac{2}{3}}(1-\Psi)^{3/2}.
\end{align}
Contrary to the bag EoS, $\alpha_{p,{\rm max}}^{\rm SS}$ does not become infinite; it is therefore always possible to find a value of $\alpha_p$ large enough to obtain a runaway solution.

\bibliographystyle{JHEP}
{\footnotesize
\bibliography{biblio}}

\end{document}